\documentclass[12pt]{article}

\usepackage{graphicx} 
\usepackage{jheppub}
\usepackage{ textcomp }
\usepackage{amsfonts,amsmath} 
\usepackage{mathrsfs} 
\usepackage{setspace}
\usepackage[utf8]{inputenc} 
\usepackage{comment} 

\makeatletter
\def\@fpheader{\relax}
\makeatother

\DeclareMathAlphabet{\mathbbold}{U}{bbold}{m}{n} 
\newcommand{\be}{\begin{equation}} \newcommand{\ee}{\end{equation}}

\DeclareMathOperator{\Tr}{Tr}

\title{Scalar quasinormal modes of Kerr-AdS$\mathbf{_5}$}

\author[a,b]{Juli\'{a}n Barrag\'{a}n Amado,}\emailAdd{j.j.barragan.amado@rug.nl}
\author[a]{Bruno Carneiro da Cunha,}\emailAdd{bcunha@df.ufpe.br}
\author[b]{and Elisabetta Pallante}\emailAdd{epallante@rug.nl}

\affiliation[a]{Departamento de F\'{i}sica, Universidade Federal de Pernambuco,
50670-901, Recife, Pernambuco, Brazil} 
\affiliation[b]{Van Swinderen Institute for Particle Physics and
  Gravity, University of Groningen, Groningen, The Nederlands} 

\abstract{An analytic expression for the scalar quasinormal modes of the
  generic, spinning Kerr-$\mathrm{AdS_5}$ black holes was previously
  proposed by the authors in \cite{Amado:2017kao}, in terms of
  transcendental equations involving the Painlevé VI (PVI) $\tau$ function. In
  this work, we carry out a numerical investigation of the modes for
  generic rotation parameters,
  comparing implementations of expansions for the PVI $\tau$ function
  both in terms of conformal blocks (Nekrasov functions) and
  Fredholm determinants. We compare the results with standard 
  numerical methods for the subcase of Schwarzschild black holes. We then
  derive asymptotic formulas for the angular eigenvalues and the quasinormal
  modes in the small black hole limit for generic scalar mass and discuss,
  both numerically and analytically, the appearance of
  superradiant modes.}


\begin{document}

\maketitle

\section{Introduction}
The quasinormal fluctuations of black holes play an important role in
general relativity (GR). Improving the precision of the quantitative
knowledge of the decay rates is required to advance our understanding of
gravitation, from the interpretation of gravitational wave data to the
study of the linear stability of a given solution to Einstein equations. 
       
A completely different motivation to analyze quasinormal oscillation of
black holes arises from the gauge/gravity correspondence. In the
context of the Maldacena's conjecture, black hole solutions in
asymptotic AdS spacetimes describe thermal states of the
corresponding CFT with the Hawking temperature, and the perturbed
black holes describe the near-equilibrium states. Namely, the
perturbation -- parametrized by a scalar field in our case of study --
induces a small deviation from the equilibrium, so that the
(scalar) quasinormal mode spectrum of the black hole is dual to poles
in the retarded Green's function on the conformal side. Thus one can
compute the relaxation times in the dual theory by equating them to the
imaginary part of the eigenfrequencies \cite{Nunez:2003eq}. There have
been many studies of quasinormal modes of various types of
perturbations on several background solutions in AdS spacetime, and we
refer to \cite{Berti:2009kk} for further discussions.  

We turn our attention to a specific background, the Kerr-$\rm AdS_5$
black hole \cite{Hawking:1998kw}. The motivation to put on a firmer
basis the linear perturbation problem of the Kerr-$\rm AdS_5$ system
is threefold. First, the calculation of scattering
coefficients/quasinormal modes depends on the connection relations of
different solutions to Fuchsian ordinary differential equations -- the
so-called connection problem, for which we present the exact solution 
in terms of transcendental equations. Second, by the AdS/CFT duality,
 perturbations of the Kerr-$\rm AdS_5$ black hole serves as a tool to
study the associated CFT thermal state \cite{Hawking:1999dp,
  Landsteiner:1999xv} with a sufficiently general set of Lorentz
charges (mass and angular momenta). Small Schwarzschild-$\rm
AdS_5$ black holes, with horizon radius smaller than the $\rm AdS$
scale, are known to be thermodynamically unstable, it would be thus
interesting to have some grasp on the generic rotating
case. Finally, numerical and analytic studies hint at the existence of
unstable (superradiant) massless scalar modes \cite{Cardoso:2004hs,
  Aliev:2008yk, Uchikata:2009zz}, which should also be well described
by the isomonodromy method. 

The Painlevé VI (PVI) $\tau$ function was introduced in this context by
\cite{Novaes:2014lha,daCunha:2015fna} -- see also
\cite{Novaes:2018fry} -- as an approach to study rotating 
black holes in four dimensions and positive cosmological constant. The
method has deep ties to integrable systems and the Riemann-Hilbert
problem in complex analysis, relating scattering coefficients to
monodromies of a flat holomorphic connection of a certain matricial
differential system associated to the Heun equation — the
isomonodromic deformations. For the Heun equation related to the
Kerr-de Sitter and Kerr-anti-de Sitter black holes, the solution for
the scattering problem has been given in terms of transcendental
equations involving the PVI $\tau$ function.  

In turn, the PVI $\tau$ function has been interpreted as a chiral
$c=1$ conformal block of Virasoro primaries, through the AGT
conjecture \cite{Gamayun:2012ma}. In the latter work, the authors have given
asymptotic expansions for the PVI $\tau$ function in terms of Nekrasov
functions, expanding early work by Jimbo \cite{Jimbo:1982}. More
recently, the authors of \cite{Gavrylenko:2016zlf,Cafasso:2017xgn} 
have re-formulated the PVI $\tau$ function in terms of the determinant of
a certain class of Fredholm operators. We will see that this formulation has
computational advantages over the Nekrasov sum expansion and will
allow us to numerically solve the transcendental equations posed by the
quasinormal modes with high accuracy.

The paper is organized as follows. In Section \ref{Kerr_AdS}, we
review the five-dimensional Kerr-AdS metric, and write the linear
scalar perturbation equation of motion in terms of the radial and the
angular Heun differential equation. In Subsection \ref{isomonodromy},
we review the isomonodromy method. First, the solutions of each Heun
equation are linked to a differential matricial differential equation,
which in turn can be seen as a flat holomorphic connection. Then, we
identify gauge transformations of each connection as a Hamiltonian 
system which is directly linked to the Painlev\'{e} VI $\tau$
function. Finally, we recast the conditions to obtain our original
differential equations and their quantization conditions in terms of
the PVI $\tau$ function. 

In Section \ref{sec3}, we give approximate expressions for the
monodromy parameters in terms of the isomonodromy time $t_0$. Applying 
these results to the angular equation, we obtain an approximate
expression for the separation constant for slow rotation or near
equally rotating black holes. We then set out to calculate numerically
the quasinormal modes for the  Schwarzschild-$\rm AdS_5$ and compare
with the established Frobenius methods and Quadratic Eigenvalue
Problem (QEP). 

In Section \ref{sec4}, we turn to the general-rotation Kerr-$\rm AdS_5$ black
holes. We study numerically the quasinormal modes for
increasing outer horizon radii, again comparing with the Frobenius
method. We then use the analytical results for the monodromy
parameters for the radial equation to give an asymptotic formula for
the quasinormal modes in the subcase where the field does not carry any
azymuthal angular momenta $m_1=m_2=0$ (and therefore the angular
eigenvalue quantum number $\ell$ even). We close by discussing the
existence of superradiant modes for $\ell$ odd.

We conclude in Section \ref{conclusions}. In Appendix A we describe
the Nekrasov expansion and the Fredholm determinant formulation of the
PVI $\tau$ function, reviewing work done in
\cite{Gavrylenko:2016zlf}. In Appendix B we give an explicit 
parametrization of the monodromy matrices given the monodromy
parameters. 

\section{Scalar Fields in Kerr-$\mathbf{AdS_5}$}
\label{Kerr_AdS}

Let us review the five dimensional Kerr-${\rm AdS}_5$ black hole
metric as presented in \cite{Hawking:1998kw}
\begin{align}
ds^{2} &=
         -\dfrac{\Delta_{r}}{\rho^{2}}\left(dt-\dfrac{a_1\sin^{2}\theta}{1-a_1^2}d\phi
         -\dfrac{a_2\cos^{2}\theta}{1-a_2^2}d\psi\right)^{2}+\dfrac{\Delta_{\theta}
         \sin^{2}\theta}{\rho^{2}}\left(a_1dt-\dfrac{(r^{2}+a_1^{2})}{1-a_1^2}d\phi
         \right)^{2} \nonumber  \\ 
&+ \dfrac{1+r^{2}}{r^{2}\rho^{2}}\left( a_1a_2dt -
  \dfrac{a_2(r^{2}+a_1^{2})\sin^{2}\theta}{1-a_1^2}d\phi -
  \dfrac{a_1(r^{2}+a_2^{2})\cos^{2}\theta}{1-a_2^2}d\psi \right)^{2}
  \nonumber \\ 
&+ \dfrac{\Delta_{\theta}\cos^{2}\theta}{\rho^{2}}\left( a_2dt-
  \dfrac{(r^{2}+a_2^{2})}{1-a_2^2}d\psi \right)^{2}
  +\dfrac{\rho^{2}}{\Delta_{r}}dr^{2}+
  \dfrac{\rho^{2}}{\Delta_{\theta}}d\theta^{2}, \label{eq1}   
\end{align}
where
\begin{gather}
\Delta_{r} =
\dfrac{1}{r^{2}}(r^{2}+a_1^{2})(r^{2}+a_2^{2})(1+r^2)-2M
=\dfrac{1}{r^2}(r^2-r_0^2)(r^2-r_-^2)(r^2-r_+^2),
             \nonumber \\ 
\Delta_{\theta} = 1- a_1^2\cos^{2}\theta -
                  a_2^{2}\sin^{2}\theta, \quad\quad \ 
\rho^{2} = r^{2} + a_1^{2}\cos^{2}\theta + a_2^{2}\sin^{2}\theta,
\label{eq2}
\end{gather}
with $M, a_1,a_2$ real parameters, related to the ADM mass and
angular momenta by \cite{Gibbons:2004ai,Hollands:2005wt,Olea:2006vd}
\begin{gather}
{\cal M}=\frac{\pi M(2\Xi_1+2\Xi_2-\Xi_1\Xi_2)}{4\Xi_1^2\Xi_2^2},\quad\quad
{\cal J}_\phi=\frac{\pi M a_1}{2\Xi_1^2\Xi_2},\quad\quad
{\cal J}_\psi=\frac{\pi M a_2}{2\Xi_1\Xi_2^2}.\\
\Xi_1=1-a_1^2,\quad\quad\quad \Xi_2=1-a_2^2. \nonumber
\end{gather}
When $M>0$, $a^2_1,a^2_2<1$ all these quantities are physically
acceptable, and we have that $r_-,r_+$, the real roots of $\Delta_r$,
correspond to the inner and outer horizons of the black hole
\cite{Gibbons:2004ai}, whereas $r_0$ is purely imaginary:
\begin{equation}
  -r_0^2=1+a_1^2+a_2^2+r_-^2+r_+^2.
\end{equation}

For the purposes of this article, we will see the radial variable, or
rather $r^2$, as a generic complex number. It will be interesting for
us to treat all three roots of $\Delta_r$, $r_+^2,r_-^2$ and $r_0^2$
as Killing horizons. Actually, in the complexified version of the
metric \eqref{eq1}, in all three hypersurfaces defined by
$r=r_0,r_-,r_+$ we have that each of the Killing fields
\begin{equation}\label{eq4}
\xi_{k} = \dfrac{\partial}{\partial t} +
\Omega_{1}(r_{k})\dfrac{\partial}{\partial \phi} +
\Omega_{2}(r_{k})\dfrac{\partial}{\partial \psi},\quad\quad k=0,-,+, 
\end{equation}
becomes null.  The temperature and angular velocities for each horizon 
are given by 
\begin{gather}
\Omega_{k,1} = \dfrac{a_1 (1-a_1^2)}{r^{2}_{k} + a_1^{2}}, \qquad
\Omega_{k,2} = \dfrac{a_2 (1-a_2^2)}{r^{2}_{k} + a_2^{2}},
\nonumber \\ 
T_{k} = \dfrac{r^{2}_{k}\Delta_r'(r_{k})}{4\pi(r^{2}_{k} +
  a_1^{2})(r^{2}_{k} + a_2^{2})}
=
\frac{r_k}{2\pi}\frac{(r_k^2-r_i^2)(r_k^2-r_j^2)}{(r_k^2+a_1^2)(r_k^2+a_2^2)},
\quad\quad i,j\neq k.
\label{eq5} 
\end{gather}
Within the physically sensible range of parameters, $T_+$ is positive,
$T_-$ is negative and $T_0$ is purely imaginary.

\subsection{Kerr-anti de Sitter scalar wave equation}
The Klein-Gordon (KG) equation for a scalar of mass $\mu$ in the
background \eqref{eq1} is separable by the factorization $\Phi =
\Pi(r)\Theta(\theta)e^{-i\omega t + im_{1}\phi +im_{2}\psi}$. To wit,
$\omega$ is the frequency of the mode, and $m_1,m_2\in\mathbb{Z}$ are
the azymuthal components of the mode's angular momentum. The angular
equation is given by \cite{Aliev:2008yk}
\begin{multline}
\dfrac{1}{\sin\theta\cos\theta}\dfrac{d}{d\theta}\left(\sin\theta
  \cos\theta\Delta_{\theta}  
  \dfrac{d\Theta(\theta)}{d\theta}\right) 
-\left[\omega^2+\frac{(1-a_1^2)m_1^2}{\sin^2\theta}+
  \frac{(1-a_2^2)m_2^2}{\cos^2\theta} \right. \\ \left.
  -\frac{(1-a_1^2)(1-a_2^2)}{\Delta_\theta}(\omega+m_1a_1+m_2a_2)^2 
+\mu^2(a_1^2\cos^2\theta+a_2^2\sin^2\theta)\right]\Theta(\theta)
=-C_j\Theta(\theta),
\label{eq8}
\end{multline}
where $C_j$ is the separation constant, and $j$ an integer index which
will be defined later. By two consecutive
transformations $\chi = \sin^{2}\theta$, and $u=\chi/(\chi -
\chi_0)$, with $\chi_0=(1-a_1^2)/(a_2^2-a_1^2)$\footnote{The second
  change of variables is justified in terms 
  of the asymptotic expansion for the $\tau$ function close to 0.
}, we can take the four singular points of \eqref{eq8} to be located at   
\begin{equation}
u=0,\quad u=1,\quad u=u_{0}=\frac{a_2^2-a_1^2}{a_2^2-1},\quad u=\infty, 
\end{equation}
and the indicial exponents\footnote{The asymptotic behavior of
the function near the singular points $\Theta(u)\simeq (u-u_i)^{\alpha^\pm_i}$
or $\Theta(u)\simeq u^{-\alpha^\pm_\infty}$ for the point at infinity.} are
\begin{gather}
\alpha^\pm_0=\pm\frac{m_1}{2},\quad
\alpha_1^\pm=\frac{1}{2}\left(2\pm\sqrt{4+\mu^2}\right),\quad
\alpha^\pm_{u_0}=\pm\frac{m_2}{2},\quad
\alpha^\pm_\infty=\pm\frac{1}{2} (\omega+a_1m_1+a_2m_2).
\end{gather}
The exponents have a sign symmetry, except for $\alpha^\pm_1$, which
corresponds $\Delta/2$ and $(4-\Delta)/2$, where $\Delta$ is the
conformal dimension of the CFT primary field associated to the
$\mathrm{AdS_5}$ scalar. We define
the single monodromy parameters $\varsigma_i$ through
$\alpha^\pm_i=\tfrac{1}{2}(\alpha_i\pm\varsigma_i)$. Writing them
explicitly 
\begin{gather}
\varsigma_0=m_1,\quad
\varsigma_1=2-\Delta,\quad
\varsigma_{u_0}=m_2,\quad
\varsigma_\infty\equiv\varsigma=\omega+a_1m_1+a_2m_2.
\label{eq:angmonodromies}
\end{gather}
We note an obvious sign symmetry $\varsigma_i\rightarrow
-\varsigma_i$, so we will take the positive sign as standard.

Coming back to the \eqref{eq8}, by introducing the following
transformation
\begin{equation}
\Theta(u) = u^{m_1/2}(u-1)^{\Delta/2}(u-u_{0})^{m_2/2}S(u)
\label{eq:shomos}
\end{equation}
we bring the angular equation to the canonical Heun form
\begin{equation}
\frac{d^2S}{du^2}+\left(\frac{1+m_1}{u}+\frac{1+\sqrt{4+\mu^2}}{u-1}
  +\frac{1+m_2}{u-u_0} \right)\frac{dS}{du}+\left(
\frac{q_1q_2}{u(u-1)}-\frac{u_0(u_0-1)Q_0}{u(u-1)(u-u_0)}\right)S=0
\label{eq:heunangular}
\end{equation}
with $q_1,q_2$ and the accessory parameter $Q_0$ given by
\begin{equation}
  q_1=\frac{1}{2}\left(m_1+m_2+\Delta-\varsigma\right),
  \quad\quad
  q_2=\frac{1}{2}\left(m_1+m_2+\Delta+\varsigma\right),
\end{equation}
\begin{multline}
4u_0(u_0-1)Q_0
=-\frac{\omega^2+a_1^2\mu^2-C_j}{a_2^2-1}-u_0\left[(m_2+
  \Delta-1)^2-m_2^2-1\right] \\ 
-(u_0-1)\left[(m_1+m_2+1)^2 -\varsigma^2 -1\right].
 \label{eq:accessoryqangular}
\end{multline}
We note that \eqref{eq:heunangular} has the same AdS spheroidal
harmonic form as the problem in four dimensions, the eigenvalues
reducing to those ones when $m_1=m_2$, $\ell\rightarrow \ell/2$,
$a_1=0$ and $a_2=i\alpha$ 
\cite{Novaes:2018fry}. Also, according to \eqref{eq8} we have that 
$u_0$ in \eqref{eq:heunangular} is close to zero for $a_2 \simeq a_1$,
the equal rotation limit.

The radial equation is given by
\begin{multline}\label{eq9}
\dfrac{1}{r\Pi(r)}\dfrac{d}{dr}\left(r\Delta_{r}
  \dfrac{d\Pi(r)}{dr}\right) - \biggl[ C_j + \mu^{2}r^{2} +
\dfrac{1}{r^{2}}(a_1a_2\omega - a_2(1-a_1^2)m_{1} - a_1(1-a_2^2)m_{2})^{2}\biggr]
+\\ 
+ \dfrac{(r^{2}+a_1^{2})^{2}(r^{2}+a_2^{2})^{2}}{r^{4}\Delta_{r}}\left(
  \omega - \dfrac{m_{1}a_1(1-a_1^2)}{r^{2}+a_1^{2}} -
  \dfrac{m_{2}a_2(1-a_2^2)}{r^{2}+a_2^{2}}\right)^{2}  = 0,
\end{multline}
which again has four regular singular points, located at the
roots of $r^2\Delta_r(r^2)$ and infinity. The indicial exponents
$\beta_i^\pm$ are defined analogously to the angular
case. Schematically, they are given by
\begin{equation}
\beta_k=\pm\frac{1}{2}\theta_k,\quad k=+,-,0\quad\text{and}\quad \quad
\beta_\infty = \frac{1}{2}(2\pm\theta_\infty), 
\end{equation}
where $\theta_k$, $k=+,-,0,\infty$ are the single monodromy
parameters, given in terms of the physical parameters of the problem by
\begin{equation}\label{eqThetas}
\theta_{k} = \dfrac{i}{2\pi}\left(\dfrac{\omega -
    m_{1}\Omega_{k,1} - m_{2}\Omega_{k,2}}{T_{k}}\right), \quad
\theta_{\infty}=2-\Delta,
\end{equation}
where $k=0,+,-$. To bring this equation to the 
canonical Heun form which we can use, we perform the change of
variables\footnote{Note that, with this choice of variables, we have
  that at infinity, the radial solution will behave as $\Pi(z) \sim
  z^{\pm\theta_{0}/2}$},  
\begin{equation}
  z=\frac{r^2-r_-^2}{r^2-r_0^2},\quad\quad
  \Pi(z) =
  z^{-\theta_{-}/2}(z-z_0)^{-\theta_{+}/2}(z-1)^{\Delta/2}R(z),
  \label{eq:shomor}
\end{equation}
where
\begin{equation}
  z_0=\frac{r_+^2-r_-^2}{r_+^2-r_0^2}.
  \label{eq:radialt0}
\end{equation}
The equation for $R(z)$ is
\begin{equation}
\dfrac{d^{2}R}{dz^{2}} + \biggl[\dfrac{1-\theta_{-}}{z}
+\dfrac{-1+\Delta}{z-1} +\dfrac{1 -\theta_{+}}{z-z_0} \biggr]\dfrac{dR}{dz} + 
\left( \frac{
    \kappa_1\kappa_2}{z(z-1)}-\frac{z_0(z_0-1)K_0}{z(z-1)(z-z_0)}\right)R(z)
=  0,
\label{eq:heunradial}
\end{equation}
where
\begin{equation}
  \kappa_1=\frac{1}{2}(\theta_{-}+\theta_{+}-\Delta-\theta_{0}),
  \quad\quad
  \kappa_2=\frac{1}{2}(\theta_{-}+\theta_{+}-\Delta+\theta_{0}),
\end{equation}
\begin{multline}
4z_0(z_0-1)K_0=-\frac{C_j+\mu^2r_{-}^2-\omega^2}{r_+^2-r_{0}^2}
-(z_0-1)[(\theta_{-}+\theta_{+}-1)^2-\theta_{0}^2-1]\\
-z_0\left[2(\theta_{+}-1)(1-\Delta)+(2-\Delta)^2-2\right]
\label{eq:accessorykradial}
\end{multline} 

Both equations \eqref{eq:heunangular} and \eqref{eq:heunradial} can be
solved by usual Frobenius methods in terms of Heun series near each of
the singular points. We are, however, interested in solutions for
\eqref{eq:heunangular} which satisfy
\begin{equation}
S(u)=
\begin{cases}
1+{\cal O}(u),& u\rightarrow 0, \\
1+{\cal O}(u-1),& u\rightarrow 1,
\end{cases}
\label{eq:boundaryfors}
\end{equation}
which will set a quantization condition for the separation constant
$C_j$. For the radial equation with $\mu^2>0$, the conditions that
$\Pi(z)$ corresponds to a purely ingoing wave at the outer horizon $z=z_{0}$
and normalizable at the boundary $z=1$ read as follows\footnote{The
  computation of the accessory parameters and the boundary conditions
  of the radial equation are slightly different with respect to those
  shown in \cite{Amado:2017kao}. We have chosen a more suitable
  M\"{o}bius transformation for the asymptotic expansion of the
  PVI $\tau$ function in the limit $z_0\rightarrow 0$.}: 
\begin{equation}
R(z)=
\begin{cases}
1+{\cal O}\left(z-z_{0}\right),& z\rightarrow z_{0}, \\
1+{\cal O}\left(z-1\right),& z\rightarrow 1, 
\end{cases}
\label{eq:boundaryforr}
\end{equation}
where $R(z)$ is a regular function at the boundaries. This
condition will enforce the quantization of the (not necessarily real)
frequencies $\omega$, which will correspond to the (quasi)-normal
modes.

\subsection{Radial and Angular $\tau$ functions}
\label{isomonodromy}
 
The functions described in this Section will
be the main ingredient to compute the separation constant $C_j$ and the
quasinormal modes, which will be the focus of next Section. A more
extensive discussion of the strategy can be found in
\cite{Amado:2017kao}. Let us begin by rewriting the Heun equation in
the standard form as a first order differential equation. Consider the
system given by 
\begin{multline}\label{eq:Fuchsian}
\frac{d\Phi}{dz}=A(z)\Phi, \quad
\Phi(z)=\begin{pmatrix}
y^{(1)}(z) & y^{(2)}(z) \\
w^{(1)}(z) & w^{(2)}(z)
\end{pmatrix}, \quad
A(z) = \frac{A_0}{z} + \frac{A_t}{z-t} +\frac{A_1}{z-1}
 \end{multline}
 where $\Phi(z)$ is a matrix of fundamental solutions and the
 coefficients $A_i$, $i=0,t,1$ are $2
 \times 2$ matrices that do not depend on $z$. Using
 \eqref{eq:Fuchsian} we can derive a second order ODE for one of the
 two linearly independent solutions $y^{(1,2)}(z)$ given by 
 \begin{equation}\label{eq:fuchsiandef}
 y^{\prime\prime} - \left( \Tr A +(\log A_{12})^{\prime}
 \right)y^{\prime} + \left( \det A - A_{11}^{\prime} + A_{11}(\log
   A_{12})^{\prime} \right)y = 0, 
\end{equation}
which, by the partial fraction expansion of $A(z)$ will have singular
points at $z=0,t,1,\infty$ and at the zeros and poles of
$A_{12}(z)$. Let us investigate the latter. By a change of basis of
solutions, we can assume that the matrix $A(z)$ becomes diagonal at
infinity, thus
\begin{align}\label{eq:Ainf}
 A_{\infty} = - \left(A_{0} + A_{1} + A_{t}\right), \quad A_{\infty}=\begin{pmatrix}
 \kappa_+ & 0 \\
 0 & \kappa_{-}
\end{pmatrix}.
\end{align}
This leads to the assumption that $A_{12}$ vanishes like
$\mathcal{O}(z^{-2})$ as $z \rightarrow \infty$. By the partial
fraction form of $A(z)$ we then have
\begin{equation}
A_{12}(z) = \frac{k(z-\lambda)}{z(z-1)(z - t)}, \quad k,\lambda \in \mathbb{C}
\end{equation}
where $k$ and $\lambda$ do not depend on $z$ but can be expressed
explicitly in terms of the entries of $A_i$, as can be seen in
\cite{Jimbo:1981-2}. For our purposes, it suffices to check that
$z=\lambda$ is a zero of $A_{12}(z)$ and necessarily of order
1. Therefore, $z=\lambda$ is an extra singular point of
\eqref{eq:fuchsiandef}, which does not correspond to the poles of
$A(z)$. A direct calculation shows that this singular point has
indicial exponents 0 and 2, with no logarithmic tails, and hence
corresponds to an apparent singularity, with trivial monodromy. The
resulting equation for \eqref{eq:fuchsiandef} is in general not quite
the Heun equation, but has five singularities
\begin{equation}
\label{eq:deformed}
\mbox{{\small$y^{\prime\prime} + \left( \frac{1-\theta_0}{z}
      +\frac{1-\theta_t}{z-t} + \frac{1-\theta_1}{z-1} -
      \frac{1}{z-\lambda} \right)y^{\prime} + \left(
      \frac{\kappa_{+}(\kappa_{-}+1)}{z(z-1)}
      -\frac{t(t-1)K}{z(z-t)(z-1)} +
      \frac{\lambda(\lambda-1)\mu}{z(z-1)(z-\lambda)} \right)y =
    0$}}. 
\end{equation}
where $\theta_i=\Tr A_{i}$ and we set by gauge transformation $\det
A_{i} =0$ for $i=0,t,1$. The accessory parameters are
$\mu=A_{11}(z=\lambda)$ and $K$, which is defined below. We
will refer to this equation as the deformed Heun equation.

The absence of logarithmic behavior at $z=\lambda$ results in the
following algebraic relation between $K$, $\mu$ and $\lambda$
\begin{equation}
K(\mu,\lambda,t) = \frac{\lambda(\lambda-1)(\lambda-t)}{t(t-1)}\left[
  \mu^2 - \left(\frac{\theta_0}{\lambda} + \frac{\theta_1}{\lambda -1}
    + \frac{\theta_t -1}{\lambda -t}\right)\mu +
  \frac{\kappa_{+}(1+\kappa_{-})}{\lambda(\lambda-1)}\right].
\label{eq:hamiltoniank}
\end{equation}
Now, since we are interested in properties of the solutions of
\eqref{eq:fuchsiandef}, and therefore of \eqref{eq:Fuchsian}, which depend
solely on the monodromy data -- phases and change of bases picked as
one continues the solutions around the singular points -- we are free
to change the parameters of the equations as long as they do not change
the monodromy data. The isomonodromy deformations parametrized by a
change of $t$ view $A(z)$ as the ``$z$-component'' of a flat
holomorphic connection ${\cal A}$. The ``$t$-component'' can be
guessed immediately:
\begin{equation}
  {\cal A}_z=A(z),\quad\quad  {\cal A}_t=-\frac{A_t}{z-t},
\end{equation}
and the flatness condition gives us the Schlesinger equations:
\begin{equation}
\begin{gathered}
  \frac{\partial A_0}{\partial t}=-\frac{1}{t}[A_0,A_t],\quad\quad
  \frac{\partial A_1}{\partial t}=-\frac{1}{t-1}[A_1,A_t],\\
  \frac{\partial A_t}{\partial
    t}=\frac{1}{t}[A_0,A_t]+\frac{1}{t-1}[A_1,A_t].
\end{gathered}
\end{equation}
When integrated, these equations will define a family of flat
holomorphic connections ${\cal A}(z,t)$ with the same monodromy data,
parametrized by a possibly complex parameter $t$. The set of
corresponding $A(z,t)$ will be called the isomonodromic family. It
has been known since the pioneering work of the Kyoto school in the
1980's -- see \cite{Iwasaki:1991} for a mathematical review and
\cite{daCunha:2015fna} for the specific case we consider here -- that
the flow defined by these equations is Hamiltonian, conveniently defined by the
$\tau$ function 
\begin{equation}
  \frac{d}{dt}\log \tau(t;\{\vec{\theta},\vec{\sigma}\})
  =\frac{1}{t}\Tr(A_0A_{t})+\frac{1}{t-1}  \Tr(A_1A_{t}).
\label{eq:thetaufunction}
\end{equation}
In terms of $\mu,\lambda$, the Schlesinger flow can be cast as
\begin{equation}
  \frac{d\lambda}{dt}=\frac{\partial K}{\partial \mu},\quad\quad
  \frac{d\mu}{dt}=-\frac{\partial K}{\partial \lambda},
  \label{eq:schlesingerhamiltonian}
\end{equation}
and the ensuing second order differential equation for $\lambda$ is
known as the PVI transcendent. The relation between the
$\tau$ function and the Hamiltonian can be obtained by direct algebra:
\begin{equation}
  \frac{d}{dt}\log \tau(t;\{\vec{\theta},\vec{\sigma}\}) =
  K(\mu,\lambda,t)+\frac{\theta_0\theta_t}{t}+\frac{\theta_1\theta_t}{t-1}-
  \frac{\lambda(\lambda-1)}{t(t-1)}\mu-\frac{\lambda-t}{t(t-1)}\kappa_+.
\end{equation}

Expansions for the
PVI $\tau$ function near $t=0,1$ and $\infty$ were given
in \cite{Gamayun:2012ma,Gamayun:2013auu}, and Appendix A. For $t$
sufficiently close to zero we have 
\begin{multline}
  \tau(t) =
  Ct^{\frac{1}{4}(\sigma^2-\theta_0^2-\theta_t^2)}(1-t)^{\frac{1}{2}\theta_1\theta_t}
  \left(1+\left(\frac{\theta_1\theta_t}{2}+
    \frac{(\theta_0^2-\theta_t^2-\sigma^2)(\theta_\infty^2-
      \theta_1^2-\sigma^2)}{8\sigma^2}\right)t\right.\\
  -\frac{(\theta_0^2-(\theta_t-\sigma)^2)
      (\theta_\infty^2-(\theta_1-\sigma)^2)}{
      16\sigma^2(1+\sigma)^2}\kappa t^{1+\sigma}\\
    \left. -\frac{(\theta_0^2-(\theta_t+\sigma)^2)
      (\theta_\infty^2-(\theta_1+\sigma)^2)}{
      16\sigma^2(1-\sigma)^2}\kappa^{-1}t^{1-\sigma}
  +\ldots\right).
\label{eq:taufunctionexpansion}
\end{multline}
The parameters in these expansions are related to the monodromy data
$\{\vec{\theta},\vec{\sigma}\}=\{\theta_0,\theta_t,\theta_1,\theta_\infty;
\sigma_{0t},\sigma_{1t}\}$, 
where $\theta_i=\Tr A_i$ are as above and $\sigma_{ij}$ are the composite
monodromy parameters
\begin{equation}
  2\cos\pi\sigma_{ij}=\Tr M_i M_j,
\end{equation}
where $M_i(M_j)$ is the matrix that implements the analytic continuation
aroud the singular point $z_i(z_j)$. Given the monodromy data, the 
$\sigma$ parameter is related to
$\sigma_{0t}$ by the addition of an even integer
$\sigma_{0t}=\sigma+2p$ so that the coefficients above will give the
largest term in the series. We will defer the procedure to calculate
$p$ until Section 4. The parameter $\kappa$ is given in terms of the
monodromy data by \eqref{eq:tildes}.

The usefulness of the PVI $\tau$ function for the solution of the
scattering and quasinormal modes for the scalar AdS perturbations
is based on the relation between the scattering coefficients and the
monodromy data \cite{Novaes:2014lha,Novaes:2018fry}. For the
quasinormal modes, the relationship was shown in
\cite{Amado:2017kao}. Succintly, it states 
that conditions like \eqref{eq:boundaryfors} and \eqref{eq:boundaryforr}
require that the relative connection matrix between the Frobenius solutions
constructed at the singular points to be upper or lower triangular. In
turn, this means that, in the basis where one monodromy matrix is
diagonal, the other will be upper or lower triangular. A direct
calculation shows that 
\begin{equation}
  \cos \pi\sigma_{ij}=\cos\pi(\theta_i+\theta_j).
  \label{eq:quasinormalmodes}
\end{equation}
As derived in \cite{Amado:2017kao}, the converse is also true: if the
composite monodromy is given by \eqref{eq:quasinormalmodes}, then the
monodromy matrices $M_i$ and $M_j$ are either both lower or upper
triangular. We note that this formulation views the problem of finding
eigenvalues for the angular equation similar in spirit to finding the
quasinormal frequencies for the radial equation. 

For the problem under consideration, the expressions for the
composite monodromies condition \eqref{eq:quasinormalmodes} in terms
of the quantities in each ODE \eqref{eq:heunangular} and
\eqref{eq:heunradial} are 
\begin{gather}
\sigma_{0u_0}(m_1,m_2,\varsigma,\Delta,u_0,C_{j})=m_1+m_2 +2j, 
\quad  \quad j \in \mathbbold{Z},
\label{eq:angularsln}\\
\sigma_{1 z_{0}}(\theta_k,\Delta,z_0,
\omega_n,C_j)=\theta_{+}+\Delta 
+2n - 2,\quad\quad
n\in \mathbbold{Z}.
\label{eq:implicitsln}
\end{gather}

These conditions on the $\tau$ function for the radial and angular
system can be obtained by first placing conditions on the matricial
system \eqref{eq:Fuchsian} such that the equation for the first line
of $\Phi(z)$ \eqref{eq:fuchsiandef} recovers the differential equation
we are considering -- \eqref{eq:heunangular} for the angular case and
\eqref{eq:heunradial} for the radial case. We need, from the generic form
of the equation satisfied by the first line \eqref{eq:deformed}, that
the canonical variables $\lambda(t_0)=t_0$, $\mu(t_0)$ and $K(t_0)$
are to be chosen so that \eqref{eq:hamiltoniank} has a
well-defined limit as $\lambda(t_0)\rightarrow t_0$. These conditions,
expressed in terms of the $\tau$ function \eqref{eq:thetaufunction}, are 
\begin{equation}
  \begin{gathered}
  \left.
      \frac{d}{dt}\log\tau(t;\{\vec{\theta},\vec{\sigma}\}^-)\right|_{t=t_0}
  =\dfrac{(\theta_{t}-1)\theta_1}{2(t_0-1)}+
  \dfrac{(\theta_t-1)\theta_{0}}{2t_0}  +K_{0}
  \\[5pt] 
  \left. \frac{d}{dt}\left[t(t-1)
      \frac{d}{dt}\log\tau(t;\{\vec{\theta},\vec{\sigma}\}^-)\right]\right|_{t=t_0} 
  =\dfrac{\theta_t-1}{2}\left(\theta_{t}-\theta_{\infty}-\theta_{0}-\theta_{1}-2\right),
  \label{eq:tauinitialconditions}
  \end{gathered}
\end{equation}
where $K_0$ is the accessory parameter of the corresponding Heun
equation (radial or angular) and the parameters of the $\tau$ function
are given by
\begin{equation}
  \{\vec{\theta},\vec{\sigma}\}^- =
  \{\theta_0,\theta_t-1,\theta_1,\theta_\infty+1;\sigma_{0t}-1,\sigma_{1t}-1\}.
  \label{eq:shiftedmonodromy}
\end{equation}
These conditions can be understood as an initial value problem of the
dynamical system defined by \eqref{eq:schlesingerhamiltonian}. Given
the expansion of the 
$\tau$ function \eqref{eq:taufunctionexpansion}, these 
conditions provide an analytic solution to the system, and can be
inverted to find the composite monodromy parameters
$\sigma_{0t},\sigma_{1t}$. We plan to apply these conditions to both
the radial equation \eqref{eq:heunradial} and the angular equation
\eqref{eq:heunangular}, and view \eqref{eq:tauinitialconditions} as
the set of (exact) transcendental equations which can be solved
numerically. 

The solution for the quasinormal modes means finding for
$\omega$, given the rest of the parameters of the differential
equations \eqref{eq:heunradial} and \eqref{eq:heunangular}, by solving
the set of four transcendental equations, the pair in the conditions
on the $\tau$ functions \eqref{eq:tauinitialconditions} for each
condition in the angular and radial equations \eqref{eq:angularsln}
and \eqref{eq:implicitsln}. The parameters for each pair are given
explicitly by
\begin{center}
\begin{tabular}{|c|c|c|c|c|c|}
 \hline 
 & $t_{0}$ & $\theta_{0}$ & $\theta_{t}$ & $\theta_{1}$ & $\theta_{\infty}$ \\  
 \hline 
 $\tau_{Rad}(t)$ & $z_0$ & $\theta_{-}$ & $\theta_{+} $ & $2-\Delta$ & $\theta_{0}$ \\ 
 \hline 
 $\tau_{Ang}(t)$ & $u_{0}$ & $-m_1$ & $-m_2$ & $2 - \Delta$ &
                                                            $\varsigma $ \\ 
 \hline 
\end{tabular} 
\end{center}

It should be noted that the conditions \eqref{eq:tauinitialconditions}
give an analytic solution for the quasinormal frequencies. The set of
transcendental (and implicit) equations is probably the best that can
be done: save for a few special cases -- see \cite{Gamayun:2013auu} --
the solution for the dynamical system \eqref{eq:schlesingerhamiltonian}
cannot be given in terms of elementary functions. On the other hand,
the true usefulness of the result \eqref{eq:tauinitialconditions}
relies on the control we have over the calculation of the PVI
$\tau$ function.

In previous work \cite{Amado:2017kao} we considered the
interpretation of the expansion \eqref{eq:taufunctionexpansion} in 
terms of conformal blocks, which in turn allow us to interpret the $\tau$
function as the generating function for the accessory parameters of
classical solutions of the Liouville differential equation -- an
important problem in the constructive theory of conformal maps
\cite{Anselmo20180080}. On the other hand, expressions like the first
equation in \eqref{eq:tauinitialconditions} could be interpreted in
the gauge/gravity correspondence as an equilibrium condition on the
angular and radial ``systems'', if one could interpret the radial
\eqref{eq:heunradial} and angular \eqref{eq:heunangular} equations as
Ward identities for different sectors in the purported boundary CFT --
see \cite{daCunha:2016crm} for comments on that direction in the
simpler case of BTZ black holes. The second condition in
\eqref{eq:tauinitialconditions} is related to an associated
$\tau$ function, with shifted monodromy arguments
\begin{equation}
  \tau(t;\{\vec{\theta},\vec{\sigma}\})\equiv
  \tau(t;\{\theta_0,\theta_t,\theta_1,\theta_\infty\},\{\sigma_{0t},\sigma_{1t}\})
\end{equation}
via the so-called ``Toda equation'' 
-- see proposition 4.2 in \cite{Okamoto:1986}, or \cite{Anselmo20180080}
for a sketch of proof. With help from the Toda equation, the second
condition in \eqref{eq:tauinitialconditions} can be more succintly phrased as
\begin{equation}
  \tau(t_0;\{\vec{\theta},\vec{\sigma}\}) = 0,
  \label{eq:asymtau}
\end{equation}
for which we will give an interpretation in terms of the Fredholm
determinant in Appendix A. In would be interesting to further
that line and explore the holographic aspects of the structure
outlined by the analytic solution, but we will leave that for future work. 

The expression for the $\tau$ function in terms of conformal blocks
\eqref{eq:taufunctionexpansion}, called Nekrasov expansion, is
suitable for the small black hole limit which we will treat
algebraically in this article. From the numerical
analysis perspective, however, it suffers from the combinatorial nature of its
coefficients -- see Appendix A, which takes exponential computational
time ${\cal O}(e^{\alpha N})$ to achieve ${\cal O}(t^N)$
precision. Because of this, we have used for the numerical analysis an
alternative formulation of the PVI $\tau$ function through Fredholm
determinants, introduced in \cite{Gavrylenko:2016zlf,Cafasso:2017xgn},
also outlined in Appendix A. This formulation achieves ${\cal O}(t^N)$
precision for the $\tau$ function in polynomial time ${\cal O}(N^\alpha)$.

\section{Painlev\'{e} VI $\tau$ function for Kerr-$\mathbf{AdS_5}$ black
  hole}  
\label{sec3}

For $u_0$ or $z_0$ sufficiently close to a critical value of the
PVI $\tau$ function ($t=0,1,\infty$), both the Nekrasov
expansion and the Fredholm determinant will converge fast. It makes
sense then to begin exploring solutions with this property. If $u_0$
is close to 0, this corresponds to the almost equally rotating
$a_1\simeq a_2$ or to the slowly rotating $a_1,a_2\simeq 0$ cases. For
$z_{0}$ close to 0 we are considering the near-extremal limit
$r_+\simeq r_-$, or small $r_+,r_-\simeq 0$ black holes.  

The procedure of solving \eqref{eq:tauinitialconditions} can be
summarized by first using the second equation to find the
parameter $s$ in the Nekrasov expansion \eqref{eq:nekrasovexpansion},
and then substituting this back in the first equation in order to
find the monodromy parameter $\sigma$ -- see \cite{Litvinov:2013sxa}
and \cite{Lencses:2017dgf}. In our application, there are some remarks
on the procedure. The first observation is that the $\tau$ function is
quasi-periodic with respect to shifts of $\sigma_{0t}$ by even integers
$\sigma_{0t}\rightarrow \sigma_{0t}+2p$: 
\begin{equation}
  \tau(t;\{\vec{\theta}\},\{\sigma_{0t}+2p,\sigma_{1t}\})
    =s^{-p}\tau(t;\{\vec{\theta}\},
    \{\sigma_{0t},\sigma_{1t}\}),\quad\quad p\in\mathbb{Z}.
    \label{eq:quasiperiodicity}
\end{equation}
This means that, upon inverting the equations \eqref{eq:angularsln}
and \eqref{eq:implicitsln}, we will obtain, rather than the
$\sigma_{0t}$ associated to the system, a parameter, which we will
call $\sigma$, related to $\sigma_{0t}$ by the shift
$\sigma_{0t}=\sigma+2p$. Let us digress over the consequences of this
periodicity by analyzing the structure of the expansion
\eqref{eq:nekrasovexpansion}. Schematically,
\begin{equation}
  \tau(t_0) = t_0^{\frac{1}{4}(\sigma^2-\theta_0^2-\theta_t^2)}
  \sum_{m\in\mathbb{Z}}P(\sigma+2m;t_0)s^mt_0^{m^2+m\sigma},
  \label{eq:schematics}
\end{equation}
where $P(\sigma+2m;t_0)$ is analytic in $t_0$, and to find the zero of
$\tau(t_0)$ as per \eqref{eq:asymtau} is useful to define
$X=st_0^\sigma$, making the expansion analytic in $t_0$ and
meromorphic in $X$. We can now solve \eqref{eq:asymtau} and 
thus define $X(\sigma,t_0)$ in terms of $\sigma$ as a series in
$t_0$. Let us classify these solutions by their leading term:
\begin{equation}
  X_p(\sigma;t_0)\equiv s_pt_0^{\sigma}=t_0^{2p+1}(x_0+x_1t_0+x_2t_0^2
  +\ldots)
\end{equation}
Depending on the sign of $\Re \sigma$, the leading term will
depend on $t_0$ or $t_0^{-1}$. We will suppose $\Re\sigma > 0$ for the
discussion. The ``fundamental'' solution $X_0$ is written as (see
\eqref{eq:tildes}) 
\begin{multline}
  X_{0}(\sigma;t_0)=\frac{\Gamma^2(1+\sigma)}{\Gamma^2(1-\sigma)}
  \frac{\Gamma(1+\tfrac{1}{2}(\theta_t+\theta_0-\sigma))
    \Gamma(1+\tfrac{1}{2}(\theta_t-\theta_0-\sigma))}{
    \Gamma(1+\tfrac{1}{2}(\theta_t+\theta_0+\sigma))
    \Gamma(1+\tfrac{1}{2}(\theta_t-\theta_0+\sigma))} \times \\
    \frac{\Gamma(1+\tfrac{1}{2}(\theta_1+\theta_\infty-\sigma))
    \Gamma(1+\tfrac{1}{2}(\theta_1-\theta_\infty-\sigma))}{
    \Gamma(1+\tfrac{1}{2}(\theta_1+\theta_\infty+\sigma))
    \Gamma(1+\tfrac{1}{2}(\theta_1-\theta_\infty+\sigma))}
  Y(\sigma;t_0)
  \label{eq:exisfromwye}
\end{multline}
with
\begin{multline}
  Y(\sigma;t_0)=
  \left[\frac{((\theta_t+\sigma)^2-\theta_0^2)
      ((\theta_1+\sigma)^2-\theta_\infty^2)}{ 16\,
  \sigma^2(\sigma- 1)^2
}t_0\right] \\ \times\left(1-
(\sigma- 1)\frac{(\theta_0^2-\theta_t^2)(\theta_1^2-\theta_\infty^2)+
    \sigma^2(\sigma- 2)^2}{2
    \sigma^2(\sigma- 2)^2} t_0+{\cal O}(t_0^2)\right)
\label{eq:kappaexpansion}
\end{multline}
Solutions with $\Re\sigma<0$ can be obtained sending $\sigma$ to
$-\sigma$ and inverting the term in square brackets in the expression
for $Y$. Solutions with higher value for $p$ will also be of
interest. These will have the leading term of order $t_0^{2p+1}$ and
can be obtained from the quasi-periodicity property
\eqref{eq:quasiperiodicity}, which translates to a shifting
property for $X_p$. From the generic structure \eqref{eq:schematics}
above we have  
\begin{equation}
\sum_{m\in\mathbb{Z}}P(\sigma+2m;t_0)X^mt_0^{m^2}
=\tilde{X}^{-p}t_0^{-p^2}\sum_{m\in\mathbb{Z}}P(\tilde{\sigma}+2m;t_0)
\tilde{X}^mt_0^{m^2}
\end{equation}
where
\begin{equation}
  \tilde{\sigma}=\sigma-2p,\quad\quad X=\tilde{X}t_0^{2p}.
\end{equation}
By this property, assuming $\Re\sigma>0$, we have that a solution
$X_p(\sigma;t_0)$ for \eqref{eq:asymtau} with leading term
of higher order in $t_0$ can be obtained from a 
fundamental solution of leading order $t_0$ with shifted $\sigma$:
\begin{equation}
  X_p(\sigma;t_0)=t_0^{2p}X_0(\sigma-2p;t_0)
  \label{eq:kappaseries}
\end{equation}
This allows us to construct a class of solutions for the conditions
\eqref{eq:tauinitialconditions} which are generic enough for our
purposes. From $X_p(\sigma;t_0)$ or $Y(\sigma;t_0)$ we can define the
parameter $\kappa$ entering the expansion
\eqref{eq:taufunctionexpansion}:
\begin{equation}
  \kappa(t_0;\{\vec{\theta},\vec{\sigma}\}) =
  Y(\sigma;t_0)t_0^{-\sigma},
  \label{eq:kappafromy}
\end{equation}
and the family of parameters $s_p$:
\begin{equation}
  s_p=X_p(\sigma;t_0)t_0^{-\sigma}=X_0(\sigma-2p;t_0)t_0^{-\sigma+2p},
  \label{eq:essseries}
\end{equation}
with $X_0$ given in terms of $Y$ as above. The knowledge of both parameters
$s_p$ and $\sigma$ is sufficient to determine the monodromy data by
\eqref{eq:ess}. 

We can now proceed to compute the accessory parameter $K_0$ in terms
of the monodromy parameter $\sigma$ by substituting $\kappa$ found
through \eqref{eq:kappafromy} back to the first equation in
\eqref{eq:tauinitialconditions}. We note that this equation has for
argument the shifted monodromy parameters
$\{\vec{\theta},\vec{\sigma}\}^-$  defined by
\eqref{eq:shiftedmonodromy}. This shift leaves the $s$ parameter
invariant
$s(\{\vec{\theta},\vec{\sigma}\}^-)=s(\{\vec{\theta},\vec{\sigma}\})$
but, because of the string of gamma functions in \eqref{eq:tildes},
the $\kappa$ parameter entering the asymptotic formula
\eqref{eq:taufunctionexpansion} will change as
\begin{equation}
  \kappa(\{\vec{\theta},\vec{\sigma}\}^-)=-
  \frac{16\,\sigma^2(\sigma-1)^2}{
    ((\theta_t+\sigma)^2-\theta_0^2)((\theta_\infty-
    \sigma+1)^2-(\theta_1+1)^2)}
  \kappa(\{\vec{\theta},\vec{\sigma}\}).
\end{equation}
Using the fundamental solution for $Y(\sigma,t_0)$
\eqref{eq:kappaexpansion} and \eqref{eq:kappafromy},
we find the first terms of the expansion of the accessory parameter
\begin{multline}
  4t_0K_0=(\sigma- 1)^2-(\theta_t+\theta_0-1)^2
  +2(\theta_1-1)(\theta_t-1)t_0 \\
    +\frac{((\sigma- 1)^2-1-\theta_0^2+\theta_t^2)
      ((\sigma-
      1)^2-1-\theta_\infty^2+\theta_1^2)}{2\,\sigma(\sigma-2)}t_0
    \\+2(\theta_1-1)(\theta_t-1)t_0^2+
    \frac{(\theta_0^2-\theta_t^2)^2(\theta_1^2-\theta_\infty^2)^2}{64} 
   \left(\frac{1}{\sigma^3}-\frac{1}{(\sigma-2)^3}\right)t_0^2\\
    -\frac{((\theta_0^2-\theta_t^2)(\theta_1^2-\theta_\infty^2)+8)^2
   -2(\theta_0^2+\theta_t^2)(\theta_1^2-\theta_\infty^2)^2
   -2(\theta_0^2-\theta_t^2)^2(\theta_1^2+\theta_\infty^2)-64}{
   32\,\sigma(\sigma-2)}t_0^2\\\
  +\frac{((\theta_0-1)^2-\theta_t^2) ((\theta_0+1)^2-\theta_t^2)
    ((\theta_1-1)^2-\theta_\infty^2)
    ((\theta_1+1)^2-\theta_\infty^2)}{32(\sigma+1)(\sigma-3)}t_0^2\\
  -\frac{1}{32}(5+14\theta_0^2-18\theta_t^2-18\theta_1^2+14\theta_\infty^2)t_0^2
  +\frac{13}{32}\sigma(\sigma-2)t_0^2+{\cal O}(t_0^3)
    \label{eq:accessoryKexpansion}
\end{multline}
for $\Re\sigma>0$. The corresponding expression for
$\Re\sigma<0$ can be obtained by sending
$\sigma\rightarrow -\sigma$. The
higher order corrections can be consistently computed from the series
derived in \cite{Lencses:2017dgf}. Note that, since any solution for
$X$ in the series \eqref{eq:kappaseries} will yield the same value for
$s$ in \eqref{eq:nekrasovexpansion}, and hence the same value for
$K_0$, the difference between $\sigma$ and $\sigma_{0t}$ is tied to
which terms of the expansion are dominant, and depends on the
particular value for $s$ and $t_0$. The generic structure of the
conformal block expansion, of which $K_0$ is the semi-classical limit,
was discussed at some length in the classical CFT literature
\cite{Zamolodchikov:1985ie,Zamolodchikov:1990ww}. The relevant facts
for our following discussion, given the generic expansion:
\begin{equation}
  4 t_0 K_0= k_0+k_1 t_0 + k_2 t_0^2+\ldots+k_n t_0^n+\ldots
  \label{eq:Kexpansiont}
\end{equation}
are as follows: $k_n$ is a rational function of the monodromy
parameters, the numerator is a polynomial in the ``external''
parameters $\theta_i$ and $\sigma$, and the denominator is a
polynomial of $\sigma$ alone; Secondly, $k_n$ is invariant under
the reflection $\sigma\leftrightarrow
2-\sigma$. Thirdly, $k_n$ has simple poles at
$\sigma=3,4,\dots,n+1$ and $\sigma=-1,-2,\ldots,-n+1$,
poles of order $2n-1$ at $\sigma=0,2$ and is analytic at
$\sigma=1$. Fourthly, the leading order term of $k_n$ near
$\sigma\simeq 2$ is (for $n\ge 1$) 
\begin{equation}
  k_n=-4\mathbf{C}_{n-1}\frac{(\theta_0^2-\theta_t^2)^n(\theta_1^2-\theta_\infty^2)^n}{
    16^n (\sigma-2)^{2n-1}}+\ldots,\quad\quad
    \mathbf{C}_n=\frac{1}{n+1}\begin{pmatrix} 2n \\ n \end{pmatrix},
    \label{eq:leadingorderkey}
\end{equation}
where $\mathbf{C}_n$ is the $n$-th Catalan number. A similar structure exists
for the fundamental solution $X_0(\sigma;t_0)$, or, rather, $Y(\sigma;t_0)$:
\begin{equation}
  Y(\sigma;t_0) = \chi_1 t_0+\chi_2 t_0^2+\ldots
\end{equation}
with leading order for each $\chi_n$ given by (for $n\ge 3$)
\begin{equation}
  \chi_n =-
  \mathbf{C}_{n-2}\frac{((\theta_t+\sigma)^2-\theta_0^2)(
   (\theta_1+\sigma)^2)-\theta_\infty^2)}{16\sigma^2(1-\sigma)^2}
  \frac{(\theta_1^2-\theta_\infty^2)^{n-1}(\theta_0^2-\theta_t^2)^{n-1}}{
  4^{n-1}\sigma^{2(n-1)}(\sigma-2)^{2(n-1)}}+\ldots,
  \label{eq:leadingorderkappa}
\end{equation}
where the implicit terms are of order ${\cal
  O}((\sigma-2)^{-2n+3})$ or higher. 

\subsection{The angular eigenvalues}

The separation constant can be calculated from the $\tau$ function
expansion by imposing the quantization condition
\eqref{eq:angularsln}. For equal rotation parameters, $a_1=a_2$, the
Heun equation reduces to a hypergeometric, and an analytic expression
in terms of finite combinations of elementary functions can be
obtained \cite{Aliev:2008yk}. We can recover the result with the
PVI $\tau$ function by taking the limit $u_0\rightarrow
0$. The leading term of \eqref{eq:accessoryKexpansion} gives the exact result
\begin{multline} \label{eq:lambda} 
C_j = \left(1-a_{1}^{2}\right)\bigl[ (m_{1}+m_{2} + 2j)(m_{1} +
m_{2}+2j - 2) - 2\omega a_1(m_{1}  + m_{2})\\ 
-a_1^2(m_{1} +m_{2})^{2}\bigr]+a_{1}^{2}\omega^{2}
+a_{1}^{2}\Delta(\Delta-4)
\end{multline}
which recovers the literature if we set the integer labelling the
angular mode as
\begin{equation}
\ell = - \left(m_{1} + m_{2} + 2j\right).
\end{equation}
We note that (some of) the $\rm SO(4)$ selection rules are encoded in the
requirement that $j$ is an integer \cite{Bander:1965rz}.

For generic angular parameters, the monodromy data of the angular equation
\eqref{eq:heunangular} is composed of the single monodromy parameters
\eqref{eq:angmonodromies}
$\{\varsigma_0,\varsigma_{u_0},\varsigma_1,\varsigma_\infty\}$ and 
the composite monodromy parameters
$\{\varsigma_{0u_0},\varsigma_{1u_0}\}$. Using 
the formula \eqref{eq:accessoryKexpansion}, the
separation constant \eqref{eq:lambda} can be written
up to third order in $u_0$ (remember that
$\varsigma=\omega+a_1m_1+a_2m_2$): 
\begin{multline} \label{eq:lambda_ell}
C_{\ell} = \omega^2+\ell(\ell+2)-\varsigma^2-\frac{a_{1}^{2}+a_{2}^{2}}{2}\left(\ell(\ell+2)-\varsigma^{2}-\Delta(\Delta-4)\right) \\
-\frac{\left(a_{1}^{2}-a_{2}^{2}\right)
  \left(m_{1}^{2}-m_{2}^{2}\right)}{2\ell(\ell+2)}
\left(\ell(\ell+2)-\varsigma^{2} 
+(\Delta-2)^2 \right) \\
-\frac{(a_{1}^{2}-a_{2}^{2})^{2}}{1-a^{2}_{2}}\bigg[\frac{\left(\ell(\ell+2)+m_{2}^2-m_1^2\right)\left(\ell(\ell+2)+(\Delta-2)^2-\varsigma^2\right)}{2\ell(\ell+2)}
\\ -\frac{13}{32}\ell(\ell+2)+\frac{1}{32}\left(5+14(m^2_{1}+\varsigma^{2})-18(m^{2}_{2}+(\Delta-2)^{2})\right) \\
-\frac{\left((m_{1}+1)^{2}-m_2^{2}\right)\left((1-m_1)^{2}-m^{2}_{2}\right)\left((\Delta-1)^{2}-\varsigma^{2}\right)\left((\Delta-3)^{2}-\varsigma^{2}\right)}{32(\ell-1)(\ell+3)}
\\
+\frac{\left((m^{2}_{1}-m^{2}_{2})((\Delta-2)^{2}-\varsigma^{2})+8\right)^{2}-64-2(m^{2}_{1}+m^{2}_{2})((\Delta-2)^{2}-\varsigma^{2})^{2}}{32\ell(\ell+2)} \\
-\frac{2(m^{2}_{1}-m^{2}_{2})^{2}((\Delta-2)^{2}+\varsigma^{2})}{32\ell(\ell+2)}
-\frac{\left(m^{2}_{1}-m^{2}_{2}\right)^{2}\left((\Delta-2)^{2}-\varsigma^{2}\right)^{2}}{64}\left(\frac{1}{(\ell+2)^{3}}-\frac{1}{\ell^{3}}\right)\bigg] \\
+ \mathcal{O}\left(\left(\frac{a_{1}^{2}-a_{2}^{2}}{1-a_{2}^2}\right)^{3}\right)
\end{multline}
This expression reduces to the ones found in \cite{Aliev:2008yk} when
$a_1\simeq a_2$. It also agrees with the expression in
\cite{Cho:2011pb} for $\Delta=4$, at least to the order given. 

With an expression for the separation constant we can address the
computation of the quasinormal modes using the two initial conditions
for the radial PVI $\tau$ function at $t_0=z_{0}$. We will
next explore this and compare with numerical results
obtained from well-established methods in numerical relativity. 

\subsection{The quasinormal modes for Schwarzschild}

In the limit $a_i \rightarrow 0$, one recovers the Schwarzschild-AdS
metric, and accordingly the radial differential equation coming from the
Klein-Gordon equation for massless scalar fields \eqref{eq9} can be
reduced to the standard form of the Heun equation. The exponents
$\theta_k$ are given by 
\begin{equation}
  \theta_+=\frac{i}{2\pi}\frac{\omega}{T},\quad
  \theta_- = 0,\quad
  \theta_0= \frac{1}{2\pi}\frac{\omega}{T}\frac{\sqrt{1+r_+^2}}{r_+},\quad
  \theta_\infty=2-\Delta,
\end{equation}
where $2\pi T=2\pi T_+=(1+2r_+^2)/r_+$ is the temperature of the black hole, given by
\eqref{eq5} by setting $a_1=a_2=r_-=0$. The mass of the black hole is
given by ${\cal M}=\frac{1}{2}r_+^2(1+r_+^2)$. We note that the system
of coordinates is different from \cite{Starinets:2002br}, and the
singular point at $r=r_+$ is mapped by \eqref{eq:shomor} to
$z_0=r_+^2/(1+2r_+^2)$.  

Likewise, the angular equation \eqref{eq8} reduces to a standard
hypergeometric form. The angular eigenvalues can be seen to be the usual
$\mathrm{SO(4)}$ Casimir: $C_{\ell} = \ell(\ell + 2)$.  In terms
of $\omega,\Delta,r_+$, the accessory parameter $K_0$ in
\eqref{eq:accessorykradial} is
\begin{equation}
  K_0 = - \frac{\omega^2}{4(1+r_+^2)} +
  \frac{1+2r_+^2}{1+r_+^2}\left[\frac{\ell(\ell+2)}{4r_+^2} +
    \frac{\Delta(\Delta-2)}{4}\right] +
  \frac{i\omega}{2r_+}\frac{1+r_+^2(2-\Delta)}{1+r_+^2}. 
\end{equation}
This, along with the quantization condition for the radial monodromies
\eqref{eq:implicitsln}, provide through
\eqref{eq:tauinitialconditions} an implicit solution for the
quasinormal modes $\omega_n$ along with the composite monodromy
$\sigma_{0t}$, as we will tackle in \ref{sec:qnm}.

In order to test the method we present in Tables \ref{tab:1} and \ref{tab:2}
the numerical solution $\omega_{n,\ell}$ for the
first quasinormal mode $n=0, \ell=0$ s-wave case, and compare with
known methods, the 
pseudo-spectral method with a Chebyshev-Gauss-Lobatto grid to solve
associated Quadratic Eigenvalue Problem (QEP), and the usual numerical
matching method based on the Frobenius expansion of the solution near
the horizon and spatial infinity\footnote{It should be noted that the
  Frobenius method is in spirit similar to the old combinatorial
  approach for the PVI $\tau$ function given by Jimbo
  \cite{Jimbo:1982}.}. The Frobenius method implements the smoothness
on the first derivative at the matching point of the two series
solutions constructed with 15 terms, at the horizon and the boundary
\cite{Molina:2010fb}. On the other hand, the pseudospectral
method relies on a grid with 120 points between 0 and 1. For a more
comprehensive reading, we recommend \cite{Dias:2015nua,
  Andrade:2017jmt}. The results for $\omega_{0,0}$ are reported in Tables
\ref{tab:1} and \ref{tab:2}. 

\begin{table}[htb]
  \centering
  \begin{tabular}{ | c | c | c | }
    \hline
    $r_{+}$ &$z_0$ & $\omega_{0,0}$  \\ \hline  
    $0.005$ & $2.49988\times 10^{-5}$ &
    $3.9998498731325748-1.5044808171834238\times 10^{-6} i$ \\ \hline
    $0.01$ & $9.99800\times 10^{-5}$ &
    $3.9993983005189682-1.2123793015872442\times 10^{-5} i$ \\ \hline
    $0.05$ & $2.48756\times 10^{-3}$ &
    $3.9844293869590734-1.7525974895168137\times 10^{-3}i$ \\ \hline
    $0.1$ & $9.80392\times 10^{-3}$ &
    $3.9355764849860639-1.7970664179740506\times 10^{-2}i$ \\ \hline
    $0.2$ & $3.70370\times 10^{-2}$ &
    $3.7906778316981978-0.1667439940917780i$ \\ \hline
    $0.4$ & $0.121212$ &
    $3.7173879743704008-0.7462495474087164i$ \\ \hline
    $0.6$ & $0.209302$ &
    $3.8914015767067012-1.3656095289384492i$ \\ \hline
  \end{tabular}
  \caption{The massless scalar field s-wave $\ell=0$ and fundamental
    $n=0$ quasinormal mode $\omega_{0,0}$ in a
    Schwarzschild-$\mathrm{AdS_5}$ background for some
    values of $r_+$. The results were obtained using the Fredholm
    determinant expansion for the $\tau$ function with $N=16$.} 
  \label{tab:1}
\end{table}

\begin{table}[htb]
\centering
\scalebox{0.75}{	
\begin{tabular}{ | c | c | c | c | c | }
\hline
$r_{+}$ & Frobenius & QEP \\ \hline  
  $0.005$ &
$3.9998498731325743-1.5044808171845522\times 10^{-6} i$ &
$3.9998483860043481-2.8895543908757586\times 10^{-5} i$ \\ \hline 
$0.01$ & 
$3.9993983005189876-1.2123793015712405\times 10^{-5} i$ &
$3.9993981402971502-2.3439366987252536\times 10^{-5} i$ \\ \hline
$0.05$ & 
$3.9844293869590911-1.7525974895155961\times 10^{-3}i$ &
$3.9844293921364538-1.7526437924554161\times 10^{-3}i$ \\ \hline
$0.1$ & 
$3.9355764849860673-1.7970664179739766\times 10^{-2}i$ &
$3.9355763694852816-1.7970671629389028\times 10^{-2}i$ \\ \hline
$0.2$ & 
$3.7906778316982394-0.1667439940917505i$ &
$3.7906771832980760-0.1667441392742093 i$ \\ \hline 
$0.4$ & 
$3.7173879743704317-0.7462495474087220i$ &
$3.7173988607936563-0.7462476412816416i$ \\ \hline 
$0.6$ & 
$3.8914015767126869-1.3656095289361863i$ &
$3.8913340701538795-1.3656086881322822i$ \\ \hline 
\end{tabular}}
\caption{The same quasinormal mode frequency $\omega_{0,0}$ computed
  using numerical matching from Frobenius solutions (with 15 terms)
  and the Quadratic Eigenvalue Problem (with 120 point-lattice).}
 \label{tab:2}
\end{table}

The Schwarzschild-AdS case has been considered before
\cite{Starinets:2002br,Nunez:2003eq,Hubeny:2009rc,Uchikata:2009zz,Uchikata:2011zz},
and should be thought of as a test of the new method. Even without
optimized code\footnote{Using Python's standard libraries for
  arbitrary precision floats. Python code for both the
  Nekrasov expansion and the Fredholm determinant can be provided upon
  request.}, the Fredholm determinant evaluation of the PVI
$\tau$ function provides a faster way of computing the normal modes
than both the numerical matching and the QEP method. Convergence is
significantly faster when compared to the other methods for small
$z_0\sim 10^{-5}$, and can provide at least 14 significant digits for
the fundamental frequencies. 

\section{Monodromy parameters for Kerr-AdS}
\label{sec4}

The fast convergence and high accuracy of the $\tau$ function calculation
is suitable for the study of small black holes. Turning our
attention to  Kerr-$\rm AdS_{5}$, we consider spinning black holes
of different angular momenta and radii. In view of
holographic applications, we make use of an extra parameter given by
the mass of the scalar field scattered by the black hole. Numerical
results are presented in Table \ref{tab:numericalkerr}\footnote{In the
  table values, we have neglected some precision in the results for
  the sake of clarity, but we can provide more accurate values upon
  request.}. 

\begin{table}[htb]
	\centering
	\scalebox{0.64}{
	\begin{tabular}{|c|c|c|c|c|c|}
	\hline 
	$r_{+}$ & $z_{0}$ & $\tau$ function & Frobenius  \\ 
	\hline
	$0.00200$&$4.0\times 10^{-8}$&$3.9999043938966996-3.9179009496192059\times 10^{-7} i$&$3.9999043938967028-3.9179009496196828\times 10^{-7} i$\\ \hline
	$0.02185$&$0.000476717$&$3.9970574292783057-1.3247529381539807\times 10^{-4} i$&$3.9970574292783089-1.3247529381539848\times 10^{-4}i$\\ \hline
	$0.06154$&$0.003758101$&$3.9760894388470440-3.4698629309308322\times 10^{-3} i$&$3.9760894388470473-3.4698629309308430\times 10^{-3} i$\\ \hline
	$0.10123$&$0.010040760$&$3.9339314599984108-1.8761575868127569\times 10^{-2} i$&$3.9339314599984140-1.8761575868127629\times 10^{-2} i$\\ \hline
	$0.14092$&$0.019098605$&$3.8762906043241960-5.7537333581688194\times 10^{-2} i$&$3.8762906043241993-5.7537333581688376\times 10^{-2} i$\\ \hline
	$0.18061$&$0.030620669$&$3.8166724096683002-1.2480348073108545\times 10^{-1} i$&$3.8166724096683035-1.2480348073108582\times 10^{-1} i$\\ \hline
	$0.22030$&$0.044236431$&$3.7668353453284391-2.1574723769724682\times 10^{-1} i$&$3.7668353453284420-2.1574723769724741\times 10^{-1} i$\\ \hline
	$0.29968$&$0.076131349$&$3.7116288122171590-4.3786490332401062\times 10^{-1} i$&$3.7116288122171622-4.3786490332401161\times 10^{-1} i$\\ \hline
	$0.37906$&$0.111610120$&$3.7104224042819611-6.8107859662243775\times 10^{-1} i$&$3.7104224042819692-6.8107859662244147\times 10^{-1} i$\\ \hline
	$0.49813$&$0.165833126$&$3.7816024214536172-1.0519267755676109 i$&$3.7816024214748239-1.0519267755684242 i$\\ \hline
	$0.61720$&$0.216209245$&$3.9134030353146323-1.4181181443831172 i$&$3.9134030400737264-1.4181181441373386 i$\\ \hline
	$0.73627$&$0.260096962$&$4.0879586460765776-1.7776344225896197 i$&$4.0879588168442726-1.7776344550831753 i$\\ \hline
	\end{tabular}}
	\caption{Fundamental modes for Kerr-$\rm AdS_{5}$,
          $\ell=m_1=m_2=0$, $a_1=0.002$,
          $a_2=0.00199$ and the mass of the scalar field is
          $7.96\times 10^{-8}$.}
        \label{tab:numericalkerr}
\end{table}

One can use the initial condition for the first derivative and
\eqref{eq:asymtau} to determine an asymptotic formula for the composite
monodromy parameters $\sigma$ and $s$ as functions of the
frequency. In the spirit of establishing the occurrence of instabilities, it is worth
looking at the small black hole limit. To better parametrize this
limit, let us define:
\begin{equation}
  a_1^2 = \epsilon_1r_+^2,\quad\quad a_2^2 = \epsilon_2r_+^2,
\end{equation}
with the understanding that $r_+^2$ is a small number. The three
parameters $r_+^2,\epsilon_1,\epsilon_2$ are sufficient to express the
other roots of $\Delta_r$ as follows
\begin{align}
  r_-^2&=\frac{1+(1+\epsilon_1+\epsilon_2)r_+^2}{2}\left(
    \sqrt{1+
    \frac{4\epsilon_1\epsilon_2r_+^2}{(1+(1+\epsilon_1+\epsilon_2)r_+^2)^2}}
  -1\right),\\
  -r_0^2&=\frac{1+(1+\epsilon_1+\epsilon_2)r_+^2}{2}\left(
    \sqrt{1+
    \frac{4\epsilon_1\epsilon_2r_+^2}{(1+(1+\epsilon_1+\epsilon_2)r_+^2)^2}}
  +1\right).
\end{align}
Since we want $r_-^2\le r_+^2$, the $\epsilon_i$ will satisfy
\begin{equation}
  \epsilon_1\epsilon_2 \le 1+(2+\epsilon_1+\epsilon_2)r_+^2\simeq 1,
\end{equation}
and we remind the reader that $\epsilon_{1,2}$ are also constrained by
the extremality condition $a_i<1$.

\begin{figure}[htb]
  \centering
  \includegraphics[width=0.75\textwidth]{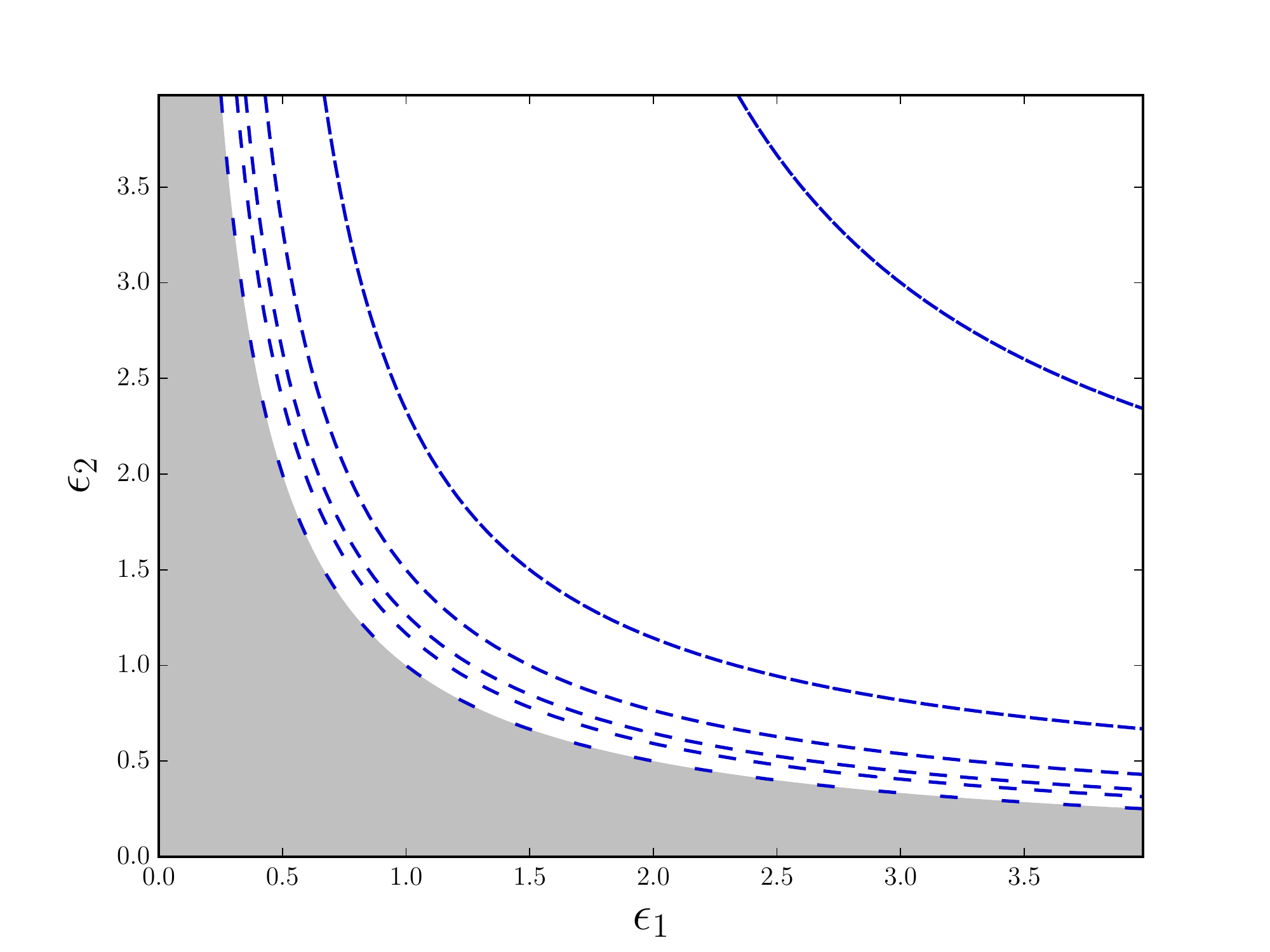}
  \caption{The space of parameters defined by $\epsilon_1\epsilon_2<1$
    corresponds to the grey area. The dashed lines represent the
    extremal black holes where $r_+=r_-$, for
    $r_+=0.002,\,0.2,\,0.25,\,0.333,\,0.5,\,1$ with increasing dash
    density. The curve $r_+=1$, the one closest to the right upper
    corner, is drawn for comparison. }
  \label{fig:2}
\end{figure}

We will focus on the case $m_1=m_2=0$ (and therefore $\ell$ even) in
order to keep the expressions reasonably short. It will be convenient to leave $z_0$ 
implicit at times,
\begin{equation}
  z_0=\frac{r_+^2-r_-^2}{r_+^2-r_0^2}=
  \frac{1+(3+\epsilon_1+\epsilon_2)r_+^2-
    \sqrt{(1+(1+\epsilon_1+\epsilon_2)r_+^2)^2+4\epsilon_1\epsilon_2r_+^2}}{
    1+(3+\epsilon_1+\epsilon_2)r_+^2+
    \sqrt{(1+(1+\epsilon_1+\epsilon_2)r_+^2)^2+4\epsilon_1\epsilon_2r_+^2}},
\end{equation}
which asymptotes as $z_0=(1-\epsilon_1\epsilon_2)r_+^2+{\cal O}(r_+^4)$.
The expansions of the single monodromy parameters
are, up to terms of order ${\cal O}(r_+^3)$:
\begin{gather}
  \theta_0=\omega\left(1-\frac{3}{2}(1+\epsilon_1)(1+\epsilon_2)
      r_+^2+\ldots\right) \label{eq:thetazeroasympt}\\ 
  \theta_+= i\omega\frac{(1+\epsilon_1)(1+\epsilon_2)}{1-\epsilon_1\epsilon_2}
  r_++\ldots\\ 
    \theta_-=-i\omega\frac{(1+\epsilon_1)
      (1+\epsilon_2)}{1-\epsilon_1\epsilon_2}\sqrt{\epsilon_1
      \epsilon_2}r_++\ldots
\end{gather}
The single monodromy parameters can be seen to have the structure:
\begin{equation}
  \theta_-=-i\phi_-r_+,\quad\quad \theta_+=i\phi_+r_+,
  \label{eq:asymptoticnotation}
\end{equation}
where $\phi_\pm$ are real and positive for real and
positive $\omega$. We also observe that $\theta_0$ is parametrically
close to the frequency $\omega$, and the correction is negative for
positive $r_+$. 

We now proceed to
solve for the composite monodromy parameter
$\sigma_\ell\equiv\sigma_{0z_0}(\ell)$
using the series expansion \eqref{eq:accessoryKexpansion}. For
even $\ell\ge 2$, the first correction is
\begin{align}
    \sigma_\ell\equiv & \ell+2 -
                        \nu_\ell\, r_+^2 \nonumber \\ 
    = &
  \ell+2
  -\frac{(1+\epsilon_1)(1+\epsilon_2)}{4(\ell+1)}(3\omega^2+3\ell(\ell+2)-
  \Delta(\Delta-4))r_+^2
  +{\cal O}(r_+^4),\quad \ell \ge 2,
  \label{eq:sigma0z0}
\end{align}
and, due to the pole structure of \eqref{eq:Kexpansiont}, naive series
inversion will yield the expansion for $\sigma$ up to order
$r_+^{2\ell}$. The case $\ell =0$ is then special, and will be dealt
with shortly. One can see from \eqref{eq:essseries} that, for $p=0$,
the monodromy parameter $s$ will behave asymptotically as
$z_0^{-\sigma}$, diverging for small $z_0$. Changing the value of $p$
will change this behavior. Changing the value of $p$ means shifting
the argument $\sigma$ that enters the definition of $X_0(\sigma,t_0)$ in
\eqref{eq:kappaseries} and therefore of $Y(\sigma,t_0)$ in
\eqref{eq:exisfromwye}. Let us call $Y_{\ell,2p}$ the expression in 
\eqref{eq:kappaexpansion} for generic $p$ and $\sigma\simeq 
2+\ell$. The expression for $p=0$ is given by 
\begin{equation}
  Y_{\ell,0}\equiv Y(\sigma_\ell;z_0) =-
  (1-\epsilon_1\epsilon_2)\frac{\omega^2-(\Delta-\ell-4)^2}{
    16(\ell+1)^2}
  \left(1+\frac{2i}{\ell+2}\phi_+r_+\right)r_+^2+\ldots,\quad \ell\ge 2. 
\end{equation}
We point out that this value is actually independent of $p$, except when
$2p=\ell$, as we will see below. We 
anticipate, from \eqref{eq:kappafromy}, that $Y_{\ell,p}$ for
$2p<\ell$ will yield a larger
value for $s_\ell$ for smaller $r_+$. We also remark that $s_\ell$ will
have a non-analytic expansion in $r_+$, due to the term
$z_0^{-\sigma_\ell}$. Finally, from the expansion we conclude that
$Y_{\ell,p}$ has an imaginary part of subleading order.

\subsection{$\ell=0$}

The ``s-wave'' case $\ell=0$ is singular since the leading behavior of
$\sigma-2$ is of order $r_+^2$. The expansion
\eqref{eq:Kexpansiont} does not 
converge in general due to the denominator structure of the
coefficients $\kappa_n$. For the small $r_+$ black hole application, however, we
are really dealing with a scaling limit where 
\begin{equation}
  \theta_-=\varphi_-\sqrt{z_0},\quad\quad \theta_+=\varphi_+\sqrt{z_0},\quad
  \text{and}\quad \sigma = 2-\upsilon z_0
\end{equation}
have finite limits for $\varphi_\pm$ and $\upsilon$ as $z_0\rightarrow
0$. Because of the poles of increasing order in $\sigma$ in
\eqref{eq:Kexpansiont}, in the $\ell=0$ case one has to resum the whole
series in order to compute $\upsilon$.

Thankfully the task is amenable due to the fact that, in the scaling
limit, the term of order $z_0$ in each of the factors $k_nt_0^n$ in
the expansion \eqref{eq:Kexpansiont} comes from the leading order pole
\eqref{eq:leadingorderkey} 
\begin{equation}
k_nz_0^n=
-4\mathbf{C}_{n-1}\frac{(\varphi_-^2-\varphi_+^2)^n(\theta_1^2-\theta_\infty^2)^n}{ 
    16^n\upsilon^{2n-1}}z_0+{\cal O}(z_0^2).
\end{equation}
The series can be resummed using the generating function for the Catalan
numbers
\begin{equation}
  1+x+2x^2+5x^3+\ldots = \sum_{n=0}^\infty
  \mathbf{C}_nx^n=\frac{1-\sqrt{1-4x}}{2x}
\end{equation}
and the result for $\upsilon$ readily written
\begin{multline}
  4z_0K_0(\ell=0)+(\theta_++\theta_--1)^2+2(\theta_1-1)(\theta_+-1)
  \frac{z_0}{z_0-1}\\
  =1+\frac{1}{2}(\theta_1^2-\theta_\infty^2)z_0-
  2\upsilon z_0\sqrt{1+
    \frac{(\varphi_+^2-\varphi_-^2)(\theta_1^2-\theta_\infty^2)}{4\upsilon^2}}
  +{\cal O}(z_0^2).
  \label{eq:upsilon0}
\end{multline}
A similar procedure allows us to compute the parameter $Y(\upsilon)\equiv
Y(2-\upsilon z_0;z_0)$ up to order $z_0^{3/2}$:
\begin{equation}
  Y(\upsilon) =
  -z_0(1+\varphi_+\sqrt{z_0})\frac{\theta_\infty^2-(\theta_1+2)^2}{64}\left(1
    +\sqrt{1+\frac{(\varphi_+^2-\varphi_-^2) (\theta_1^2-
        \theta_\infty^2)}{4\upsilon^2}}\right)^2+\ldots
  \label{eq:kappa0}
\end{equation}

For the application to the $\ell=0$ case of the scalar field, we will
use the notation \eqref{eq:asymptoticnotation}, and again use
$\sigma_0=2-\nu_0r_+^2$. In terms of the black hole pameters, $\nu_0$
has a surprisingly simple form:
\begin{equation}
  \nu_0=\frac{1}{4}(1+\epsilon_1)(1+\epsilon_2)
  \sqrt{(3\omega^2-\Delta(\Delta-4))^2-4\omega^2(\omega^2-(\Delta-2)^2)}
  +{\cal O}(r_+^2),
\end{equation}
and
\begin{multline}
  Y_{0,0}\equiv Y(\sigma_0;z_0)=-(1-\epsilon_1\epsilon_2)r_+^2
  \left(1+i\phi_+ r_+\right) \frac{\omega^2-(\Delta-4)^2}{64}\\
   \times\left(1+\frac{3\omega^2-\Delta(\Delta-4)}{
    \sqrt{(3\omega^2-\Delta(\Delta-4))^2-4\omega^2(\omega^2-(\Delta-2)^2)}}
    \right)^2 +\ldots 
\end{multline}

Finally, let us define the shifted $Y_{\ell,\ell}$ for
$2p=\ell$. Since the shifted argument $\sigma-2p$ is close to $2$,
we need the same scaling  limit as above in \eqref{eq:kappa0}. The result is
  \begin{multline}
  Y_{\ell,\ell}\equiv Y(\sigma_\ell-\ell;z_0)=-(1-\epsilon_1\epsilon_2)r_+^2
    \left(1+i\phi_+ r_+\right) 
   \frac{\omega^2-(\Delta-4)^2}{64} \\
   \times\left(1+\sqrt{1+\frac{4(\ell+1)^2\omega^2
         (\omega^2-(\Delta-2)^2)}{(3\omega^2+3\ell(\ell+2)-
         \Delta(\Delta-4))^2}}\right)^2+\ldots,
\end{multline}
where $\nu_\ell$ is taken from \eqref{eq:sigma0z0}.

To sum up, we exhibit the overall structure for small $r_+$:
\begin{gather}
  \sigma_\ell=\ell+2-\nu_\ell r_+^2+\ldots,\\
  Y_{\ell,\ell} = -
  (1-\epsilon_1\epsilon_2)\vartheta_\ell\left(1+i\phi_+r_+\right)r_+^2+\ldots,
  \label{eq:sigmakappaell}
\end{gather}
where $\nu_\ell$ and $\vartheta_\ell$ have non-zero limits as
$r_+\rightarrow 0$, corrections of order $r_+^2$, and, most
importantly, are positive for $\omega$ real and
greater than $\Delta-4$.

\subsection{The quasinormal modes}
\label{sec:qnm}

Implementation of the quantization condition \eqref{eq:implicitsln}
can be done with the formula \eqref{eq:sigma1t}. This yields a
transcendental equation for $\omega$ whose solutions will 
give all complex frequencies for the radial quantization
condition. These include negative-real part frequencies, as well as
non-normalizable modes. Since we are interested in positive 
real-part frequencies, we will consider a small correction to the
vacuum $\mathrm{AdS}_5$ result \cite{Aharony:1999ti,Penedones:2016voo} 
\begin{equation}
  \omega_{n,\ell}=\Delta+2n+\ell +\eta_{n,\ell}\, r_+^2,
  \label{eq:omegacorrection}
\end{equation}
under the hypothesis that $\eta_{n,\ell}$ has a finite limit as
$r_+\rightarrow 0$. One notes by \eqref{eq:thetazeroasympt} that
$\theta_0$ and $\omega$ are perturbatively close, so $\eta_{n,\ell}$
can be calculated perturbatively from the expansion of $\theta_0$. We
will assume that $\Delta$ is \textit{not} an integer.

The parametrization \eqref{eq:omegacorrection} allows us to expand
\eqref{eq:essseries} as a function of $r_+$. The procedure is
straightforward: We use $Y_{\ell,0}$ from
\eqref{eq:sigmakappaell}, as it gives the right asymptotic
behavior, to compute $X_0$ using \eqref{eq:exisfromwye} and then
the $s$ parameter \eqref{eq:essseries}. To second order in
$r_+$ we have  
\begin{multline}
  s_{n,\ell} = -\frac{16\,\Gamma(n+\ell/2+1)\Gamma(\Delta-2+n+\ell/2)}{
    \Gamma(n+\ell/2+3)\Gamma(\Delta+n+\ell/2)}
  \frac{\nu_\ell^2}{
    (1-\epsilon_1\epsilon_2)^2(\phi_+^2-\phi_-^2)}\\ \times 
  \left(1-i\phi_+r_++2i
    \frac{\phi_+\nu_\ell r_+}{\phi_+^2-\phi_-^2}\right) 
    Y_{\ell,\ell}r_+^{-2+2\nu_\ell r_+^2},
\end{multline}
and the leading behavior for the parameter $s_{n,\ell}$ given
$Y_{\ell,\ell}$ in \eqref{eq:sigmakappaell} is
\begin{equation}
  s_{n,\ell}=\Sigma_{n,\ell}
  \left(1+\frac{2i\nu_{n,\ell} r_+}{(1+\epsilon_1)(1+\epsilon_2)
    (\Delta+2n+\ell)}\right)
  r_+^{2\nu_{n,\ell}r_+^2}+\ldots,
  \label{eq:asymptoticess}
\end{equation}
where we defined $\nu_{n,\ell}$ as the correction for $\sigma$ as in
\eqref{eq:sigmakappaell} calculated at the vacuum frequency
$\nu_\ell(\omega=\Delta+2n+\ell)$. Finally,
\begin{equation}
  \Sigma_{n,\ell}=\frac{16\,\Gamma(n+\ell/2+1)\Gamma(\Delta-2+n+\ell/2)}{
    \Gamma(n+\ell/2+3)\Gamma(\Delta+n+\ell/2)}
  \frac{\nu_{n,\ell}^2 \,\vartheta_{n,\ell}}{
    (1+\epsilon_1)^2(1+\epsilon_2)^2(\Delta+2n+\ell)^2},
\end{equation}
again, with $\vartheta_{n,\ell}=\vartheta(\omega=\Delta+2n+\ell)$.
We also note that $\Sigma_{n,\ell}$ is real and positive under the
same conditions as \eqref{eq:sigmakappaell}. Moreover, the
choice of $p$, implicit in $Y_{\ell,\ell}$ guarantees that $s_{n,\ell}$
has a finite limit as $r_+\rightarrow 0$, although its dependence on
$r_+$ is non-analytic. 

\begin{figure}[htb]
  \centering
  \includegraphics[width=0.49\textwidth]{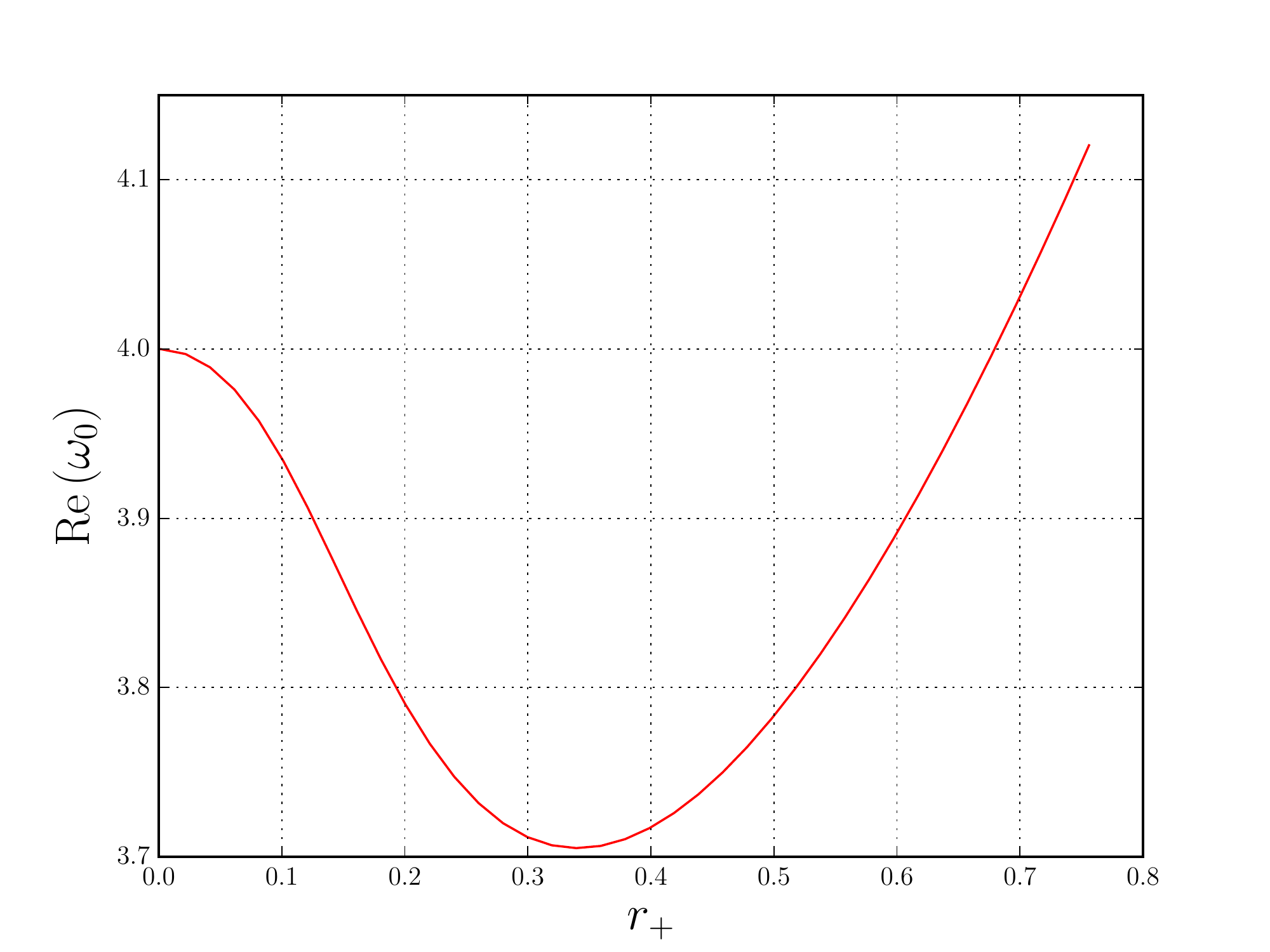}
  \includegraphics[width=0.49\textwidth]{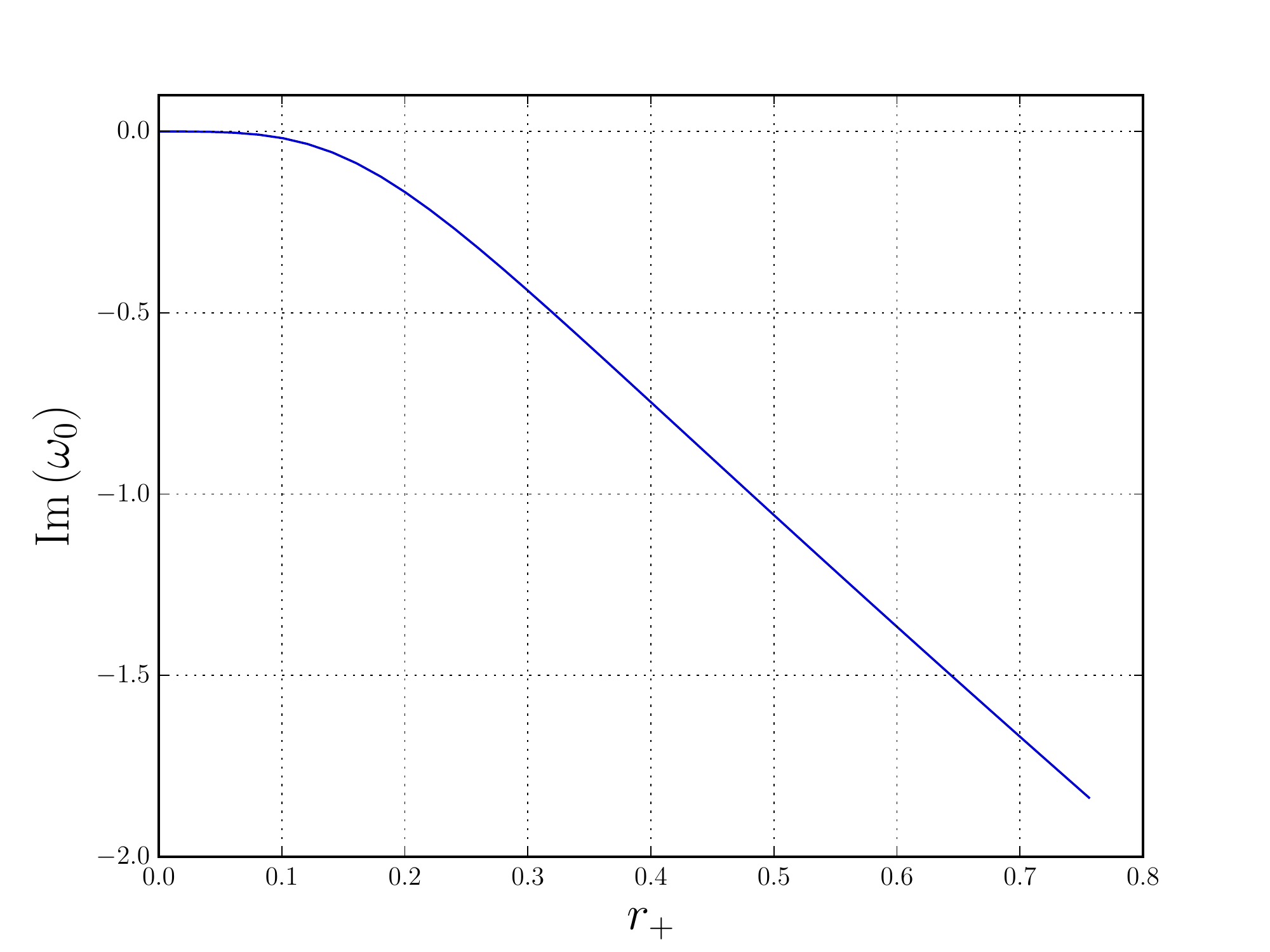}
  \includegraphics[width=0.49\textwidth]{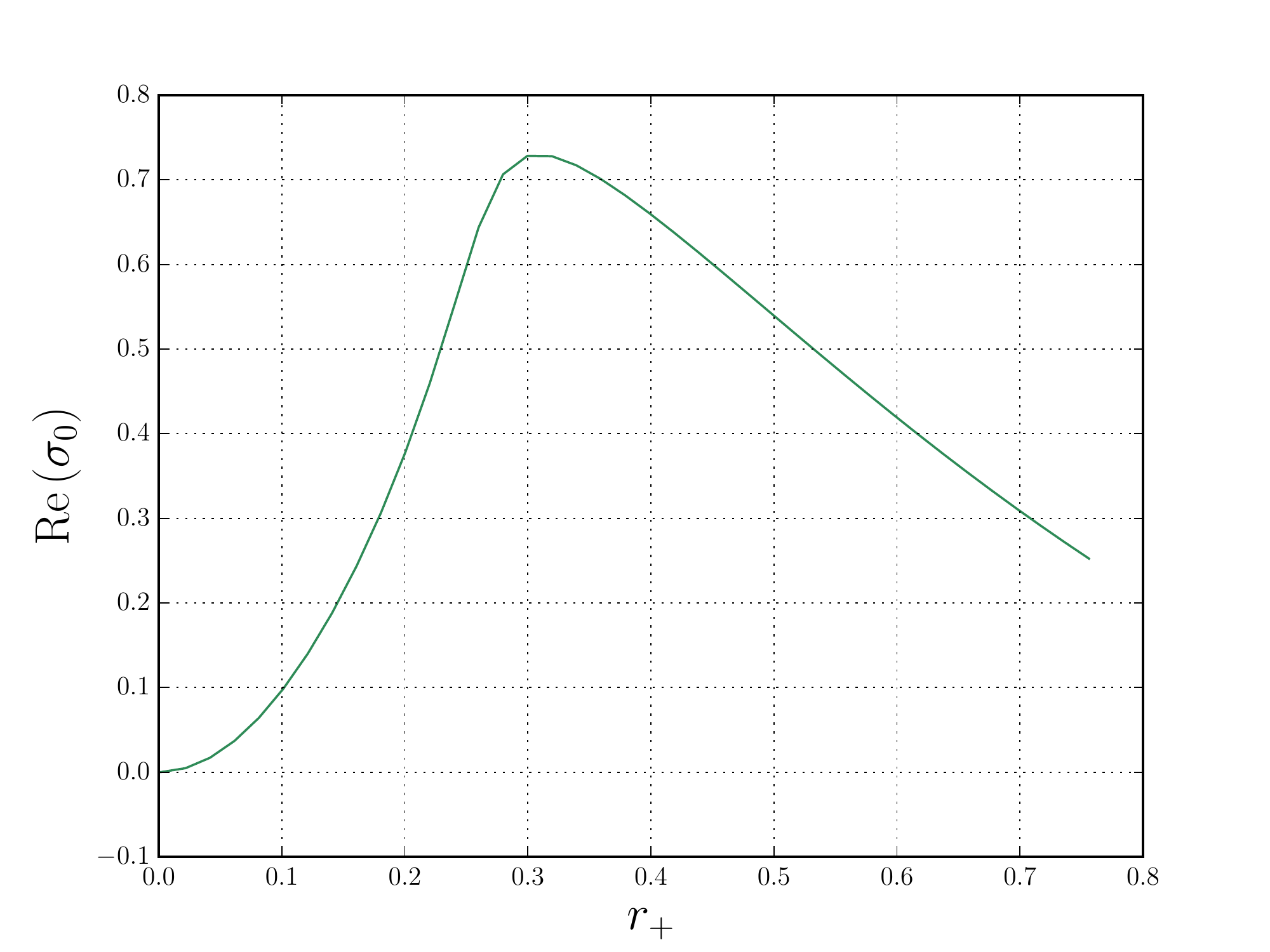}
  \includegraphics[width=0.49\textwidth]{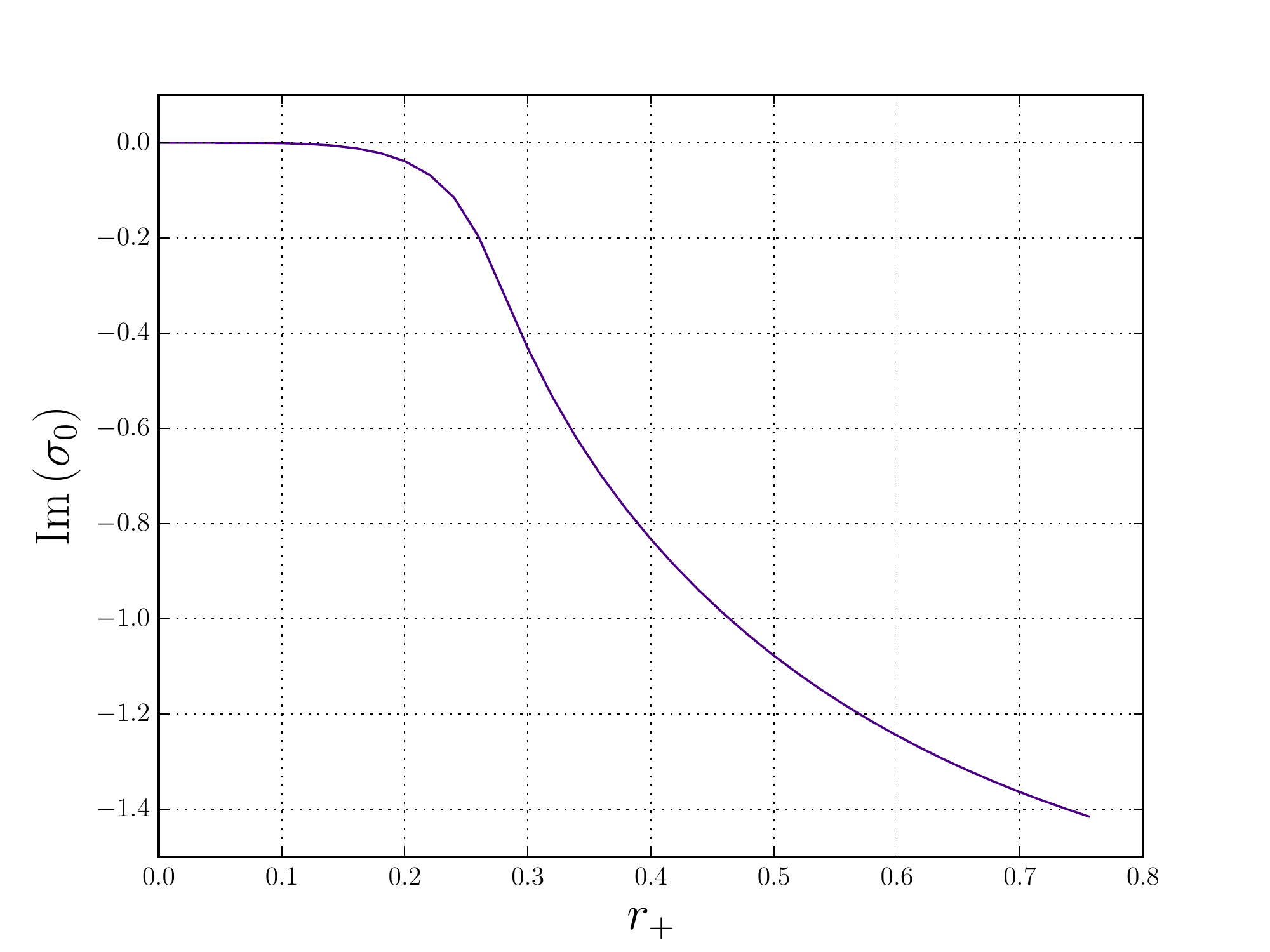}
  \caption{In the first row, the dependence of the real and
    imaginary parts of the first quasinormal mode frequency
    $\omega_0$ for small Kerr-AdS$_5$ black holes
    ($a_1=0.002,a_2=0.00199,\mu=7.96\times 10^{-8}$ ). In the
    second, the dependence of the composite parameter $2-\sigma$.}
  \label{fig:1}
\end{figure}

Equation \eqref{eq:sigma1t} can now be used, setting
$\cos\pi\sigma_{1t}=\cos\pi(\theta_1+\theta_t)$ for the radial
parameters, to find a perturbative equation for $\eta_{n,\ell}$. We
expand each of the terms in \eqref{eq:sigma1t} using
\eqref{eq:asymptoticnotation} as well as 
\begin{equation}
  \theta_0=\omega_0-\beta r_+^2,\quad
  \omega_0=\Delta+2n+\ell,\quad
  \text{and}\quad\sigma= 2+\ell-\nu_\ell r_+^2.
\end{equation}
Now, the following two relations hold
 \begin{multline}
\sin^2\pi\sigma\cos\pi(\theta_1+\theta_t) -
  \cos\pi\theta_0\cos\pi\theta_\infty-
  \cos\pi\theta_t\cos\pi\theta_1\\
  +\cos\pi\sigma(\cos\pi\theta_0\cos\pi\theta_1+
  \cos\pi\theta_t\cos\pi\theta_\infty) = \\
  \frac{\pi^3}{2}\sin(\pi\Delta)(\phi_+^2-\phi_-^2)
  \left(\beta+
    \frac{2i\nu_\ell^2r_+}{(1+\epsilon_1)(1+\epsilon_2)\omega_0}\right) r_+^4
  +\ldots
  \label{eq:sigma1trhs}
\end{multline}
\begin{multline}
    -\frac{1}{2}(\cos\pi\theta_\infty-\cos\pi(\theta_1\pm\sigma))
    (\cos\pi\theta_0-\cos\pi(\theta_t\pm\sigma)) = \\
    \frac{\pi^3}{2}\sin(\pi\Delta)(\phi_+^2-\phi_-^2)
    \left(\frac{\beta\pm\nu}{2}\right) \left(1\pm \frac{2i\nu_\ell
        r_+}{(1+\epsilon_1)(1+\epsilon_2)\omega_0} \right)
    r_+^4+\ldots
    \label{eq:sigma1tlhs}
\end{multline}

We can now proceed to calculate the first correction to the
eigenfrequencies \eqref{eq:omegacorrection}. By using the
approximations \eqref{eq:sigma1trhs} and \eqref{eq:sigma1tlhs} above
we find the correction to $\theta_0$ for each of the modes $n,\ell$
\begin{equation}
  \beta_{n,\ell}=\nu_{n,\ell}\frac{\Sigma_{n,\ell}+1}{\Sigma_{n,\ell}-1}+
  4i\frac{\nu_{n,\ell}^2}{(1+\epsilon_1)(1+\epsilon_2)(\Delta+2n+\ell)}
  \frac{\Sigma_{n,\ell}}{(\Sigma_{n,\ell}-1)^2}r_+
  +{\cal O}(r_+^2\log r_+).
\end{equation}
Finally, after some laborious calculations,
we find
\begin{multline}
  \eta_{n,\ell}=-\frac{(1+\epsilon_1)(1+\epsilon_2)}{2}
  \left[\frac{Z_{n,\ell}}{2(\ell+1)}- 3(\Delta+2n+\ell)\right]  \\
  -\frac{i}{4}(2n+\ell+1)(1+\epsilon_1)(1+\epsilon_2)
  (\Delta+2n+\ell)(2\Delta+2n+\ell-2)r_+\\
  +{\cal O}(r_+^2\log r_+),\quad\quad \ell\ge 2,
  \label{eq:etaellge2}
\end{multline}
with
\begin{multline}
  Z_{n,\ell}^2=(3(\Delta+2n+\ell)^2+3\ell(\ell+2)-
      \Delta(\Delta-4))^2\\ +4(\ell+1)^2(\Delta+2n+\ell)^2
      (2n+\ell+1)(2\Delta+2n+\ell-2).
\end{multline}    
For $\ell=0$, the form of the correction is slightly
different. Repeating the calculation, we see that $\eta_{n,\ell=0}$
has the simpler form 
\begin{multline}
  \eta_{n,0}=-\frac{(1+\epsilon_1)(1+\epsilon_2)}{4}
   (3(\Delta+2n-1)^2-(\Delta-2)^2+1) \\
   -i(n+1)(1+\epsilon_1)(1+\epsilon_2)(\Delta+2n)
   (\Delta+n-1)r_++{\cal O}( r_+^2\log r_+).
    \label{eq:etaswave}
\end{multline}
We note that both the real and
imaginary parts of the corrections $\eta_{n,\ell}$ are negative, the
real part of order $r_+^2$ as expected, and the imaginary part of order
$r_+^3$. We stress that we are taking $m_1=m_2=0$.

In the midst of the calculation, we
see that the imaginary part of $\eta_{n,0}$ has the same sign as the
imaginary part of $\theta_+$, which in turn is essentially the entropy
intake of the black hole as it absorbs a quantum of frequency $\omega$
and angular momenta $m_1$ and $m_2$:
\begin{equation}
  \theta_+=\frac{i}{2\pi}\delta S=\frac{i}{2\pi}\frac{\omega -
    m_1\Omega_{+,1}-m_2\Omega_{+,2}}{T_+}
\end{equation}
giving the same sort of window for unstable modes parameters $m_1,m_2$
as in superradiance, so a closer look at higher values for $m_{1,2}$ is
perhaps in order for future work. A full consideration of linear
perturbations of the five-dimensional Kerr-AdS black hole, involving
higher spin \cite{Wu:2009ug,Lunin:2017drx}, can be done within the
same theoretical framework presented here, and will be left for the
future. We close by observing that the expressions
\eqref{eq:etaellge2} and \eqref{eq:etaswave} above seem to represent a
distinct limit than the results in \cite{Aliev:2008yk} -- which are, however,
restricted to $\Delta=4$ -- and therefore not allowing for a direct comparison.

\subsection{Some words about the $\ell$ odd case}

Let us illustrate the parameters for the subcase $m_1=\ell,m_2=0$. The
single monodromy parameters admit the expansion
\begin{gather}
  \theta_0=\omega+\sqrt{\epsilon_1}\ell
  r_+-\frac{3}{2}(1+\epsilon_1)(1+\epsilon_2)\omega r_+^2+\ldots \\
  \theta_+=-i\ell\frac{\sqrt{\epsilon_1}(1+\epsilon_2)}{1-\epsilon_1
    \epsilon_2}+i\omega\frac{(1+\epsilon_1)(1+\epsilon_2)}{1-\epsilon_1
    \epsilon_2}r_++\ldots\\
  \theta_-=i\ell\frac{\sqrt{\epsilon_2}(1+\epsilon_1)}{1-\epsilon_1
    \epsilon_2} -i\omega
  \frac{(1+\epsilon_1)(1+\epsilon_2)}{1-\epsilon_1
    \epsilon_2}\sqrt{\epsilon_1 \epsilon_2}r_++\ldots,
\end{gather}
with all of them finite and non-zero as $r_+\rightarrow 0$. As
usual, $\theta_\pm$ are purely imaginary whereas $\theta_0$ is real
for real $\omega$. These properties hold for any value of $m_1$ and
$m_2$. 

For $\ell\ge 1$ and odd, the composite monodromy parameters are found
much in the same way as the case $\ell \ge 2$ considered above, by
inverting \eqref{eq:accessoryKexpansion}. In the following we set
$\omega_0=\Delta+2n+\ell$ as the limit of the frequency as 
$r_+\rightarrow 0$. We have for the composite monodromy parameter 
\begin{equation}
  \sigma_\ell = 2+\ell -\nu_\ell r_+^2+{\cal O}(r_+^4),
\end{equation}
with $\nu_\ell$, defined as in \eqref{eq:asymptoticess}, now for $\ell > 1$
\begin{equation}
  \nu_\ell  = (1+\epsilon_1)(1+\epsilon_2)
  \frac{3\omega_0^2+3\ell(\ell+2)-\Delta(\Delta-4)}{4(\ell+1)}
  ,\quad\quad
  \ell\ge 3.
\end{equation}
For $\ell=1$, finding $\nu_1$ from condition
\eqref{eq:tauinitialconditions} requires going to higher order in
$z_0$, due to the pole at $\sigma=3$ in the expansion
\eqref{eq:accessoryKexpansion},
\begin{multline}
    \nu_1  = \frac{(1+\epsilon_1)(1+\epsilon_2)}{32}
    (3\omega_0^2+9-\Delta(\Delta-4))  
    \\ \times\left(2 
      +\frac{1}{3}\sqrt{34-8\frac{2\Delta^4-16\Delta^3+
          (50-3\omega_0^2)\Delta^2+12(\omega_0^2-6)\Delta -36\omega_0^2}{
          (3\omega_0^2+9-\Delta(\Delta-4))^2}}\right). 
\end{multline}

For the following discussion, we take from this calculation that the
$\nu_\ell$'s are real and greater than $1$ for $\Delta > 1$, which we
will assume to hold for any $m_1$ and $m_2$. Apart from these
properties, the particular form for $\nu_\ell$ will be left
implicit. Given $\nu_\ell$, we can 
use the same procedure as in the even $\ell$ case to compute the $s$
parameter. Again, in order to have a finite $r_+\rightarrow 0$ limit,
we take $p=(\ell+1)/2$. After some calculations, we have
\begin{equation}
s_{n,\ell}=1+2\nu_\ell r_+^2\log r_++\Xi_{n,\ell}\nu_\ell r_+^2+{\cal
  O}(r_+^4(\log r_+)^2)
\label{eq:essellodd}
\end{equation}
with
\begin{multline}
\Xi_{n,\ell}= 2\gamma + \log(1-\epsilon_1\epsilon_2)+
  \Psi\left(\frac{1+\theta_++\theta_-}{2}\right)+
  \Psi\left(\frac{1+\theta_+-\theta_-}{2}\right)\\+
  \Psi\left(\frac{3+2n+\ell}{2}\right)+
  \Psi\left(\frac{3-2\Delta-2n-\ell}{2}\right),
\end{multline}
where $\Psi(z)$ is the digamma function, and $\gamma=-\Psi(1)$ the
Euler-Mascheroni constant. In the definition above we have already set
$\theta_0=\Delta+2n+\ell-\beta_{n,\ell} r_+^2$,
but as we can see from \eqref{eq:essellodd}, now we need $s_{n,\ell}$
to second order in the expansion 
parameter $r_+$. We again assume that $\Delta$ is in general
\textit{not} an integer, 
since this is irrelevant for the determination of the imaginary part
of the frequency. However, having $\Delta$ integer will change the
behavior of the real part of the correction to the eigenfrequency with
respect to $r_+$.

We note that $s_{n,\ell}$ is non-analytic, and therefore the expansion
for $\beta_{n,\ell}$ will include terms like $\log r_+$. We expand
\eqref{eq:sigma1t}, with $\sigma_{1t}=2-\Delta+\theta_+$ (up to an
even integer) to fourth order and find as first approximation to the
correction to the frequency
\begin{equation}
  \eta_{n,\ell} = \ldots +\nu_{n,\ell}
  \left(\frac{\nu_{n,\ell}+1}{\nu_{n,\ell}-1}+\Xi_{n,\ell}\right)
  \frac{r_+^2}{\log r_+}+\ldots,
  \label{eq:etaellodd}
\end{equation}
where the terms left out are real, stemming from the relation between
$\theta_0$ and $\omega$.

\begin{figure}[htb]
  \centering
  \includegraphics[width=0.495\textwidth]{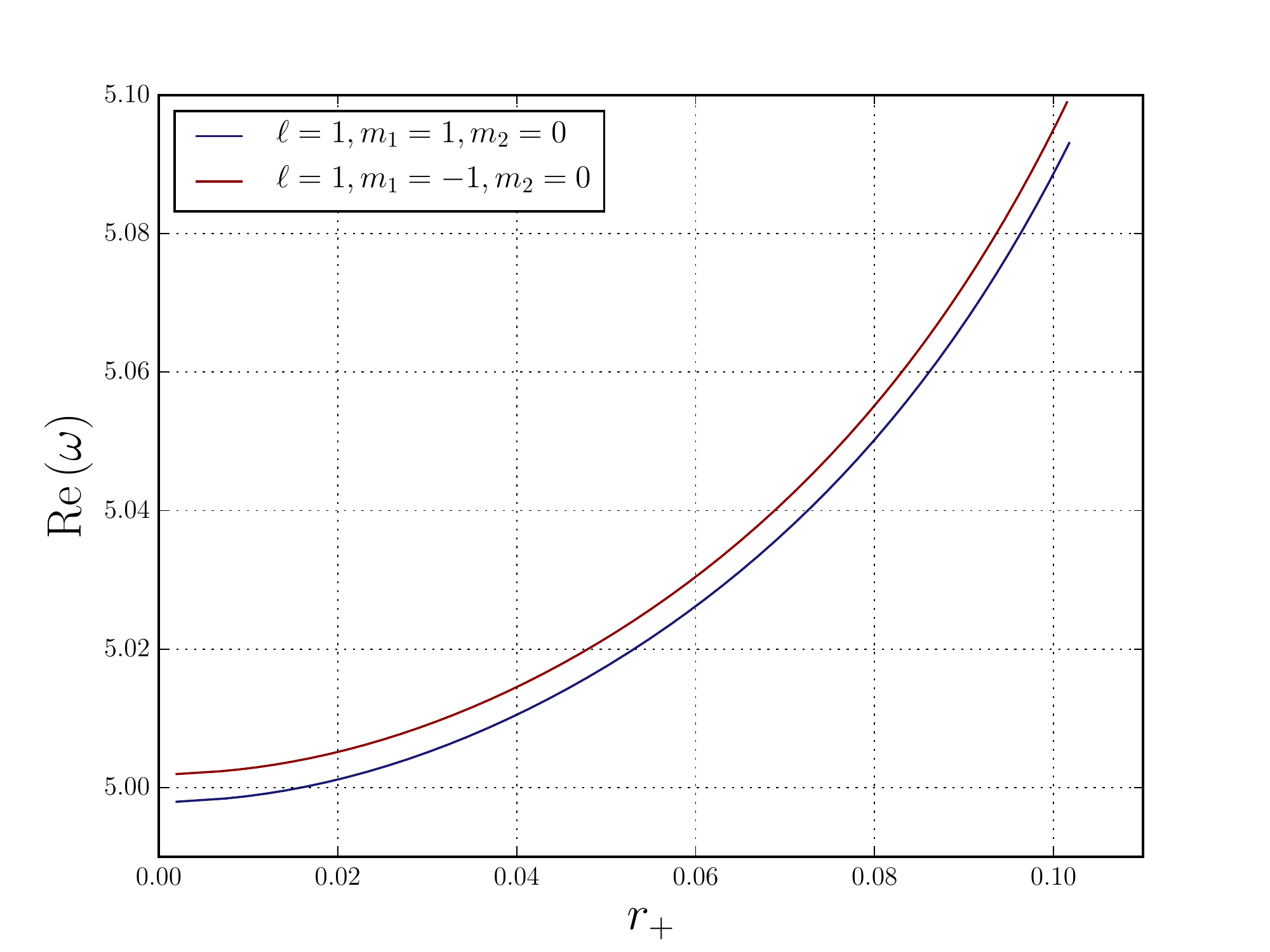}
  \includegraphics[width=0.495\textwidth]{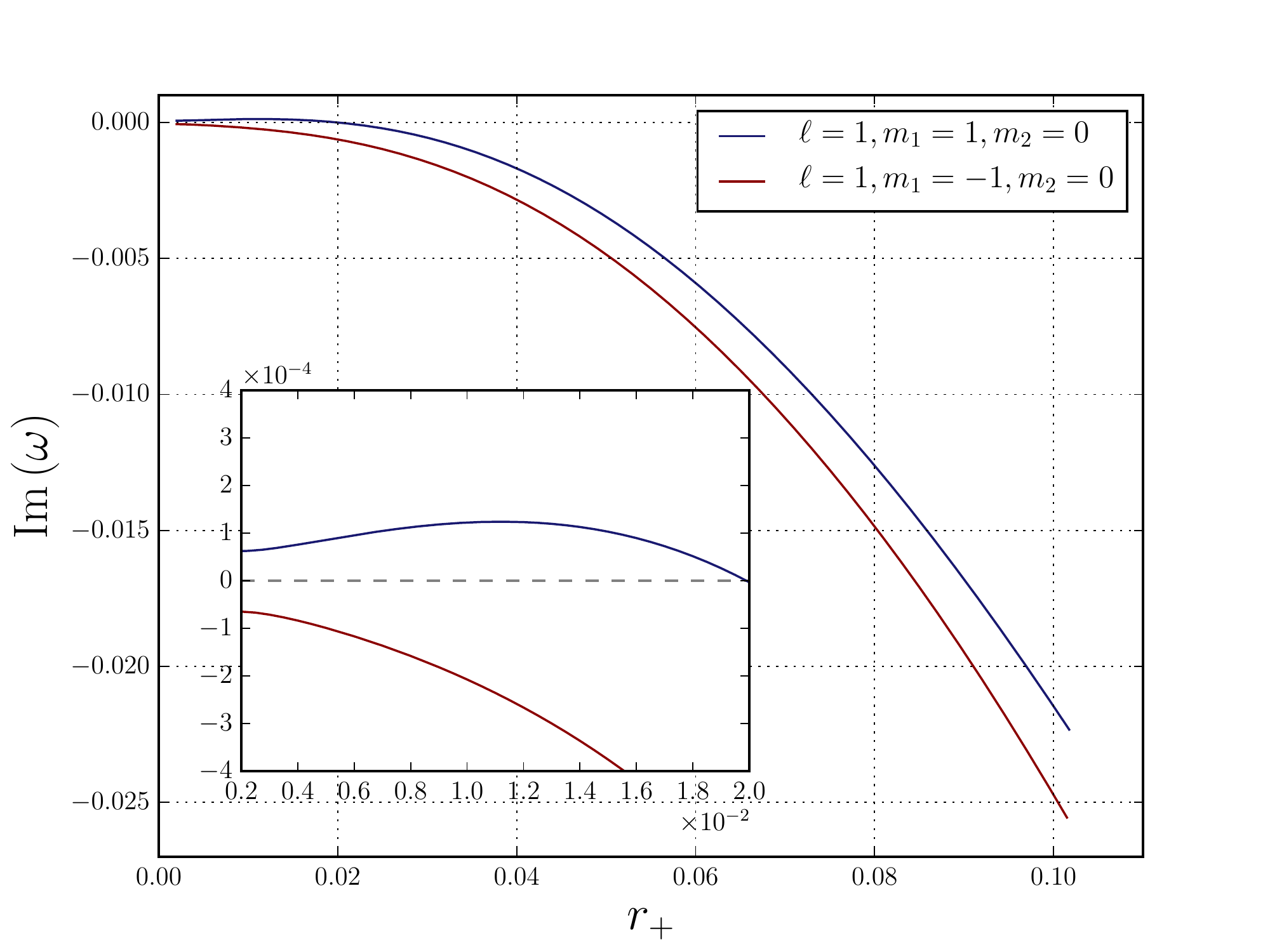}
  \caption{The dependence of the real (left) and
    imaginary part (right) of the first quasinormal mode frequency
    $\omega_0$ at $\ell=1$ and $m_1=\pm 1$ for small Kerr-AdS$_5$ black holes
    ($a_1=0.002,a_2=0.00199,\mu=7.96\times 10^{-8}$ ).}
  \label{fig:3}
\end{figure}

From \eqref{eq:etaellodd}, any possible imaginary part for the
eigenfrequency will then come from 
the imaginary part of $\Xi_{n,\ell}$. The latter can be calculated by
using the reflexion property of the digamma function
\begin{equation}
  \Im \Xi_{n,\ell}=-\frac{i\pi}{2}\left(\tan\frac{\pi}{2}(\theta_++\theta_-)
    +\tan\frac{\pi}{2}(\theta_+-\theta_-)\right),
\end{equation}
or, in terms of $m_1$ and $m_2$,
\begin{equation}
  \Im \Xi_{n,\ell}=\frac{\pi}{2}\tanh\left(\frac{\pi}{2}\frac{\sqrt{\epsilon_1}
      -\sqrt{\epsilon_2}}{1+\sqrt{\epsilon_1\epsilon_2}}(m_1-m_2)\right)
  +
\frac{\pi}{2}\tanh\left(\frac{\pi}{2}\frac{\sqrt{\epsilon_1}
    +\sqrt{\epsilon_2}}{1-\sqrt{\epsilon_1\epsilon_2}}(m_1+m_2)\right).
\label{eq:impartellodd}
\end{equation}
We then see that the imaginary part of $\Xi_{n,\ell}$ can have any sign, a
strong indication that the $\ell$ odd modes are unstable. Numerical
support for this is included in Fig. \ref{fig:3}, in which we use an
arbitrary-precision Python code (capped at 50 decimal places) to show
a slightly positive imaginary part for the resonant frequency at
$r\lesssim 0.02$. We point out that, indeed, instabilities in asymptotically
anti-de Sitter spaces are expected from general grounds
\cite{Green:2015kur}, and odd $\ell$ instabilities for the massless
case ($\Delta = 4$) were found in \cite{Aliev:2008yk}. 

\section{Discussion}

In this paper we used the isomonodromy method to derive asymptotic
expressions for the separation constant for the angular equation
(angular eigenvalue) in \eqref{eq:lambda_ell} as well as the
frequencies for the scalar quasinormal modes in a five-dimensional
Kerr-AdS background in the limit of small black holes, see in
particular \eqref{eq:etaellge2} and \eqref{eq:etaswave}. The numerical
analysis carried out for the Schwarzschild-AdS and Kerr-AdS cases
showed that the $\tau$ function approach has advantages when compared
to standard methods, in terms of faster processing times. For $\ell$
even, the correction to the vacuum AdS frequencies is negative with
negative imaginary part for $\Delta>1$, the scalar unitarity bound,
showing no instability in the range studied. For $\ell$ odd, there are
strong indications for instability due to the general structure of the
corrections in \eqref{eq:impartellodd}. In particular, for $\ell=1$,
the numerical results shown in Fig. \ref{fig:3} exhibit an unstable
mode for $r_+\leq 0.02$ and nearly equal rotational parameters. We
plan to address the phase space of instabilities and holographic
consequences in future work. 
 
The method in this paper relies on the construction of the $\tau$
function of the PVI transcendent proposed in the literature following
the AGT conjecture. The conditions in \eqref{eq:tauinitialconditions}
translate the accessory parameters in the ODEs governing the
propagation of the field -- themselves depending on the physical
parameters -- into monodromy parameters, and the quantization
condition \eqref{eq:angularsln} allows us to derive
the angular separation constant \eqref{eq:lambda_ell}. In turn, the
quantization condition for the radial equation \eqref{eq:implicitsln},
through series solutions for the composite monodromy parameters
$s$ and $\sigma$, allows us to solve for the
eigenfrequencies $\omega_{n,\ell}$, even in the generic complex
case.

The interpretation of the ODEs involved as the level 2 null
vector condition of semi-classical Liouville field theory allows us to
conclude that all descendants are relevant for the calculation of the
monodromy parameters, even though for angular momentum parameter
$\ell\ge2$, one can consider just the conformal primary (first
channel) for the parameter $\sigma_{0t}$.

The scaling limit resulting
from this analysis gives the monodromy parameter $\sigma$ in
\eqref{eq:sigma0z0}. For the parameter
$\sigma_{1t}$, the requisite of a smooth $r_+\rightarrow 0$ limit
forces us to consider the asymptotics of the whole series
\eqref{eq:accessoryKexpansion}, thus involving all descendants. This means
that naive matching of the solution obtained from the near horizon
approximation to the asymptotic solution near infinity is not a
suitable tool for dealing with small black holes. For the composite
monodromy parameter $\sigma_{1t}$, more suitably parametrized 
by $s$ in \eqref{eq:sigma1t}, the requirement of a finite
$r_+\rightarrow 0$ limit allows us to select the solutions
\eqref{eq:asymptoticess} for $\ell$ even and \eqref{eq:essellodd} for
$\ell$ odd. Although finite in the small black hole limit, the $s$
parameter has a non-analytic expansion
in terms of $r_+^m(\log r_+)^n$.

For the s-wave $\ell=0$ calculations, we had to
consider a scaling limit in \eqref{eq:accessoryKexpansion} where the
Liouville momenta associated to  $\theta_+$ and $\theta_-$ go to zero
as $r_+$, at the same time as $z_0$ and $\sigma-2$ scales as
$r_+^2$. The formulas \eqref{eq:upsilon0} and \eqref{eq:kappa0} are
reminiscent of the light-light-heavy-heavy limit of Witten diagrams
for conformal blocks \cite{Hijano:2015zsa}. It would be interesting to
understand the CFT meaning of this limit. 

The Toda equation, which allows us to interpret the second condition
\eqref{eq:tauinitialconditions} on the Painlev\'e $\tau$ function,
also merits further study. As for the first condition, we note that
it provides the accessory parameters for both the angular and radial
equations --  $Q_0$ in \eqref{eq:heunangular} and $K_0$ in
\eqref{eq:heunradial}, respectively -- as the derivative of the
logarithm of the $\tau$ function for each system. On the other hand,
these accessory parameters are both related to the separation constant
of the Klein-Gordon equation, as can be verified through
\eqref{eq:accessoryqangular} and
\eqref{eq:accessorykradial}. Including these terms in the definition  
of a $\tau$ function for the angular and radial systems, we can
represent the fact that the separation constant is the same for
\eqref{eq:heunangular} and \eqref{eq:heunradial} as the condition
\begin{equation}
  \frac{d}{du_0}\log\tau_{\mathrm{angular}}=
  \frac{d}{dz_0}\log\tau_{\mathrm{radial}},
\end{equation}
which in turn can be interpreted as a thermodynamical equilibrium
condition. Given the usual interpretation of the $\tau$ function as
the generating  functional of a quantum theory, the elucidation of
this structure can shed light on the spacetime approach to conformal
blocks. The present work gives, in our opinion, convincing evidence
that the PVI $\tau$ function is the best tool -- both numerically and
analytically -- to study connection problems for Fuchsian equations,
in particular scattering and resonance problems for a wide class of
black holes. 

\label{conclusions}
\section*{Acknowledgments}

The authors are greatly thankful to Tiago Anselmo, Rhodri Nelson,
Fábio Novaes and Oleg Lisovyy for discussions, ideas and
suggestions. We also thank Vitor Cardoso for suggestions on an early
version of this manuscript. BCdC is thankful to PROPESQ/UFPE, CNPq and
FACEPE for partial support under grant no. APQ-0051-1.05/15.

\appendix

\section{Nekrasov expansion and Fredholm determinant for Painlevé VI}

In what follows, we will assume the ``sufficient generality condition''
\begin{equation}
  \sigma_{0t}\notin\mathbb{Z},\quad
  \sigma_{0t}\pm\theta_0\pm\theta_t\notin\mathbb{Z},\quad
  \sigma_{0t}\pm\theta_1\pm\theta_\infty\notin\mathbb{Z}.
  \label{eq:sufficientlygeneral}
\end{equation}

The Nekrasov expansion of the PVI $\tau$ function is given as
a double expansion \cite{Gamayun:2013auu,Anselmo20180080}
\begin{equation}
  \tau(t)=\sum_{n\in\mathbb{Z}} {\cal
    N}^{\theta_1}_{\theta_\infty,\sigma_{0t}+2n}
  {\cal
    N}^{\theta_t}_{\sigma_{0t}+2n,\theta_0}s^nt^{\tfrac{1}{4}((\sigma_{0t}+2n)^2-
    \theta_0^2-\theta_t^2)}(1-t)^{\tfrac{1}{2}\theta_1\theta_t}
  \sum_{\lambda,\mu\in \mathbb{Y}}{\cal
    B}_{\lambda,\mu}(\vec{\theta},\sigma_{0t}+2n) t^{|\lambda|+|\mu|},
\label{eq:nekrasovexpansion}
\end{equation}
where
\begin{equation}
  {\cal N}^{\theta_3}_{\theta_2,\theta_1}=\frac{\prod_{\epsilon=\pm}
    G(1+\tfrac{1}{2}( \theta_3+\epsilon(\theta_2+\theta_1)))
    G(1+\tfrac{1}{2}( \theta_3+\epsilon(\theta_2-\theta_1)))}{
    G(1-\theta_1)G(1+\theta_2)G(1-\theta_3)}
\end{equation}
with $G(z)$ the Barnes function, defined by the solution of the
functional equation $G(z+1)=\Gamma(z)G(z)$, with $G(1)=1$ and
$\Gamma(z)$ the Euler gamma\footnote{Since $\tau$ is defined up to a
  multiplicative constant, this functional relation is only property
  of the Barnes function necessary for obtaining the expansion.}. The
other parameters in \eqref{eq:nekrasovexpansion} are the coefficients
of the $c=1$ Virasoro conformal block
\begin{multline}
  {\cal B}_{\lambda,\mu}(\vec{\theta},\sigma)=
  \prod_{(i,j)\in\lambda}\frac{((\theta_t+\sigma+2(i-j))^2-\theta_0^2)
    ((\theta_1+\sigma+2(i-j))^2-\theta_\infty^2)}{16h_\lambda^2(i,j)
    (\lambda'_j-i+\mu_i-j+1+\sigma)^2} \times \\
  \prod_{(i,j)\in\mu}\frac{((\theta_t-\sigma+2(i-j))^2-\theta_0^2)
    ((\theta_1-\sigma+2(i-j))^2-\theta_\infty^2)}{16h_\lambda^2(i,j)
    (\mu'_j-i+\lambda_i-j+1-\sigma)^2},
\end{multline}
where $\mathbb{Y}$ denote the space of Young diagrams, $\lambda$ and
$\mu$ are two of its elements, with number of boxes $|\lambda|$ and
$|\mu|$. For each box situated at $(i,j)$ in $\lambda$, $\lambda_i$ are
the number of boxes at row $i$ of $\lambda$, $\lambda'_j$, the number
of boxes at column $j$ of $\lambda$;
$h(i,j)=\lambda_i+\lambda'_j-i-j+1$ is the hook length of the box at
$(i,j)$. Finally, the parameter $s$ is given in
terms of monodromy data by: 
\begin{equation}
  s=\frac{(w_{1t}-2p_{1t}-p_{0t}p_{01}) - (w_{01}-2p_{01}-p_{0t}p_{1t})
    \exp({\pi i \sigma_{0t}})}{(2\cos \pi
    (\theta_t-\sigma_{0t})-p_0)(2\cos \pi
    (\theta_1-\sigma_{0t})-p_{\infty})}
  \label{eq:ess}
\end{equation}
where
\begin{equation}
\begin{gathered}
  p_i=2\cos\pi\theta_i,\quad  p_{ij}=2\cos\pi\sigma_{ij}, \\
  w_{0t}=p_0p_t+p_1p_{\infty},\quad w_{1t}=p_1p_t+p_0p_{\infty},\quad
  w_{01}=p_0p_1+p_tp_{\infty}.
\end{gathered}
\end{equation}

The Fredholm determinant representation for the PVI $\tau$ function uses
the usual Riemann-Hilbert problem formulation in terms of Plemelj
(projection) operators and jump matrices. The idea is to introduce
projection operators which act on the space of (pair of) functions on the
complex plane to give analytic functions with prescribed monodromy
(Cauchy-Riemann operators). Details can be found in
\cite{Gavrylenko:2016zlf}. One should point out that the two
expansions agree as functions of $t$ up to a multiplicative constant.
\begin{equation}
  \tau(t)=\mathrm{const.}\cdot
  t^{\frac{1}{4}(\sigma^2-\theta_0^2-\theta_t^2)}
  (1-t)^{-\frac{1}{2}\theta_t\theta_1}\det(\mathbbold{1}-AD),
\label{eq:fredholmexpansion}
\end{equation}
where the Plemelj operators $A,D$ act on the space of pairs of 
square-integrable functions defined on ${\cal C}$, a circle on the
complex plane with radius $R<1$:
\begin{equation}
  (Ag)(z)=\oint_{\cal C} \frac{dz'}{2\pi i}A(z,z')g(z'),\quad
  (Dg)(z)=\oint_{\cal C} \frac{dz'}{2\pi i}D(z,z')g(z'),\quad
  g(z')=\begin{pmatrix}
    f_+(z) \\
    f_-(z)
  \end{pmatrix}
  \label{eq:fredholmad}
\end{equation}
with kernels given, for $|t|<R$, explicitly by 
\begin{equation}
  \begin{gathered}
    A(z,z')=\frac{\Psi(\theta_1,\theta_\infty,\sigma;
      z)\Psi^{-1}(\theta_1,\theta_\infty,\sigma;z')-\mathbbold{1}}{z-z'},\\ 
    D(z,z')=\Phi(t)\frac{\mathbbold{1}-\Psi(\theta_t,\theta_0,-\sigma;t/z)
      \Psi^{-1}(\theta_t,\theta_0,-\sigma;t/z')}{z-z'}\Phi^{-1}(t).
  \end{gathered}
\end{equation}
The parametrix $\Psi(z)$ and the  ``gluing'' matrix $\Phi(t)$ are
\begin{equation}
  \resizebox{.90\hsize}{!}{$\displaystyle
    \Psi(\alpha_1,\alpha_2,\alpha_3;z) = \begin{pmatrix}
      \phi(\alpha_1,\alpha_2,\alpha_3;z) &
      \chi(\alpha_1,\alpha_2,\alpha_3;z) \\
      \chi(\alpha_1,\alpha_2,-\alpha_3;z) &
      \phi(\alpha_1,\alpha_2,-\alpha_3,z)
    \end{pmatrix},\quad
    \Phi(t)=\begin{pmatrix}
      t^{-\sigma/2}\kappa^{-1/2} & 0 \\
      0 & t^{\sigma/2}\kappa^{1/2}
    \end{pmatrix}$},
\end{equation}
with $\phi$ and $\chi$ given in terms of Gauss' hypergeometric function:
\begin{equation}
  \resizebox{.95\hsize}{!}{$\displaystyle 
  \begin{gathered}
    \phi(\alpha_1,\alpha_2,\alpha_3;z) = {_2F_1}(
    \tfrac{1}{2}(\alpha_1+\alpha_2+\alpha_3),\tfrac{1}{2}(\alpha_1-\alpha_2+\alpha_3);
    \alpha_3;z) \\
    \chi(\alpha_1,\alpha_2,\alpha_3;z) =
    \frac{\alpha_2^2-(\alpha_1+\alpha_3)^2}{4\alpha_3(1+\alpha_3)}
      z\,{_2F_1}(
      1+\tfrac{1}{2}(\alpha_1+\alpha_2+\alpha_3),
      1+\tfrac{1}{2}(\alpha_1-\alpha_2+\alpha_3);
      2+\alpha_3;z).
    \end{gathered}
     $}
\end{equation}
Finally, $\kappa$ is a known function of the monodromy
parameters:
\begin{multline}
  \kappa=s\frac{\Gamma^2(1-\sigma)}{\Gamma^2(1+\sigma)}
  \frac{\Gamma(1+\tfrac{1}{2}(\theta_t+\theta_0+\sigma))
    \Gamma(1+\tfrac{1}{2}(\theta_t-\theta_0+\sigma))}{
    \Gamma(1+\tfrac{1}{2}(\theta_t+\theta_0-\sigma))
    \Gamma(1+\tfrac{1}{2}(\theta_t-\theta_0-\sigma))} \times \\
    \frac{\Gamma(1+\tfrac{1}{2}(\theta_1+\theta_\infty+\sigma))
    \Gamma(1+\tfrac{1}{2}(\theta_1-\theta_\infty+\sigma))}{
    \Gamma(1+\tfrac{1}{2}(\theta_1+\theta_\infty-\sigma))
    \Gamma(1+\tfrac{1}{2}(\theta_1-\theta_\infty-\sigma))} .
\label{eq:tildes}
\end{multline}

Meaningful limits for integer $\sigma$, violating
\eqref{eq:sufficientlygeneral} can be obtained by cancelling the
factors in the denominator of $s$ with poles of the Barnes function
from the structure constants ${\cal
  N}^{\sigma+2n}_{\theta_t,\theta_0}$.   

For the numerical implementation, we write the matrix elements of $A$
and $D$ in the Fourier basis $z^n$, truncated up to order $N$. Again,
the structure of the matrix elements $A_{mn}$ and $D_{mn}$ can be
found in \cite{Gavrylenko:2016zlf}.  This truncation gives $\tau$ up
to terms ${\cal O}(t^N)$, and, unlike the Nekrasov expansion, can be
computed in polynomial time. The formulation does in principle allow for
calculation for arbitrary values of $t$, by evaluating the integrals in
\eqref{eq:fredholmad} as Riemann sums using quadratures
\cite{Bornemann:2009aa}, so there are good perspectives for using the
method outlined here for more generic configurations.

\section{Explicit monodromy calculations}

Given $\sigma_{0t}$ and $s$, satisfying \eqref{eq:sufficientlygeneral}
we can construct an explicit representation for the monodromy matrices
-- up to conjugation -- as follows. 

The monodromy matrices are
\begin{equation}
 \resizebox{.90\hsize}{!}{$\displaystyle M_0 =\frac{i}{\sin\pi\sigma_{0t}}
  \begin{pmatrix}
  \cos \pi\theta_t - \cos\pi\theta_0 e^{i\pi\sigma_{0t}} &
   2s_i\sin\frac{\pi}{2}(\sigma_{0t}+\theta_0-\theta_t)
    \sin\frac{\pi}{2}(\sigma_{0t}-\theta_0-\theta_t) \\
   -2s_i^{-1}\sin\frac{\pi}{2}(\sigma_{0t}+\theta_0+\theta_t)
    \sin\frac{\pi}{2}(\sigma_{0t}-\theta_0+\theta_t) &
    -\cos\pi\theta_t+\cos\pi\theta_0 e^{-i\pi\sigma_{0t}} 
  \end{pmatrix}$}
\end{equation}
\begin{equation}
   \resizebox{.90\hsize}{!}{$\displaystyle M_t = \frac{i}{\sin\pi\sigma_{0t}}
  \begin{pmatrix}
    \cos \pi\theta_0 - \cos\pi\theta_t e^{i\pi\sigma_{0t}} &
   -2s_ie^{i\pi\sigma_{0t}}\sin\frac{\pi}{2}(\sigma_{0t}+\theta_0-\theta_t)
    \sin\frac{\pi}{2}(\sigma_{0t}-\theta_0-\theta_t) \\
    2s_i^{-1}e^{-i\pi\sigma_{0t}}\sin\frac{\pi}{2}(\sigma_{0t}+\theta_0+\theta_t)
    \sin\frac{\pi}{2}(\sigma_{0t}-\theta_0+\theta_t)
    & -\cos\pi\theta_0+\cos\pi\theta_t e^{-i\pi\sigma_{0t}}
  \end{pmatrix}$}
  \label{eq:matrixmt}
\end{equation}
\begin{equation}
   \resizebox{.90\hsize}{!}{$\displaystyle M_1  = \frac{i}{\sin\pi\sigma_{0t}}
  \begin{pmatrix}
   -\cos \pi\theta_\infty + \cos\pi\theta_1 e^{-i\pi\sigma_{0t}} &
    2s_ee^{i\pi\sigma_{0t}}\sin\frac{\pi}{2}(\sigma_{0t}+\theta_1+\theta_\infty)
    \sin\frac{\pi}{2}(\sigma_{0t}+\theta_1-\theta_\infty) \\
    -2s_e^{-1}e^{-i\pi\sigma_{0t}}\sin\frac{\pi}{2}(\sigma_{0t}-\theta_1+\theta_\infty)
    \sin\frac{\pi}{2}(\sigma_{0t}-\theta_1-\theta_\infty) &
    \cos\pi\theta_\infty-\cos\pi\theta_1 e^{i\pi\sigma_{0t}}
  \end{pmatrix}$}
  \label{eq:matrixm1}
\end{equation}
\begin{equation}
   \resizebox{.90\hsize}{!}{$\displaystyle M_\infty =\frac{i}{\sin\pi\sigma_{0t}}
  \begin{pmatrix}
    -\cos \pi\theta_1 + \cos\pi\theta_\infty e^{-i\pi\sigma_{0t}} &
    -2s_e\sin\frac{\pi}{2}(\sigma_{0t}+\theta_1+\theta_\infty)
    \sin\frac{\pi}{2}(\sigma_{0t}+\theta_1-\theta_\infty) \\
   2s_e^{-1}\sin\frac{\pi}{2}(\sigma_{0t}-\theta_1+\theta_\infty)
    \sin\frac{\pi}{2}(\sigma_{0t}-\theta_1-\theta_\infty)
    & \cos\pi\theta_1-\cos\pi\theta_\infty e^{i\pi\sigma_{0t}}
  \end{pmatrix}$}
\end{equation}
The matrices satisfy
\begin{equation}
  M_tM_0=\begin{pmatrix}
    e^{i\pi\sigma_{0t}} & 0 \\
    0 & e^{-i\pi\sigma_{0t}}
  \end{pmatrix},\quad\quad
  M_\infty M_1=\begin{pmatrix}
    e^{-i\pi\sigma_{0t}} & 0 \\
    0 & e^{i\pi\sigma_{0t}}
  \end{pmatrix}.
\end{equation}
The parameters $s_i,s_e$ are related to the parameter
$s$ defined in \eqref{eq:ess} through:
\begin{equation}
  s=\frac{s_i}{s_e}.
\end{equation}
A direct calculation shows that
\begin{multline}
  \sin^2\pi\sigma_{0t}\cos\pi\sigma_{1t} =
  \cos\pi\theta_0\cos\pi\theta_\infty+
  \cos\pi\theta_t\cos\pi\theta_1\\
  -\cos\pi\sigma_{0t}(\cos\pi\theta_0\cos\pi\theta_1+
  \cos\pi\theta_t\cos\pi\theta_\infty)\\
  -\frac{1}{2}(\cos\pi\theta_\infty-\cos\pi(\theta_1-\sigma_{0t}))
  (\cos\pi\theta_0-\cos\pi(\theta_t-\sigma_{0t}))s\\  
  -\frac{1}{2}(\cos\pi\theta_\infty-\cos\pi(\theta_1+\sigma_{0t}))
  (\cos\pi\theta_0-\cos\pi(\theta_t+\sigma_{0t}))
  s^{-1}  
  \label{eq:sigma1t}
\end{multline}
We close by noting that for the special case of interest where
$\sigma_{1t}=\theta_1+\theta_t+2n$, $n\in\mathbb{Z}$, the expressions
above are still valid.


\providecommand{\href}[2]{#2}\begingroup\raggedright\endgroup

\end{document}